\documentclass[10pt,preprint2]{aastex}
\usepackage{color}
\usepackage{url}
\usepackage{bm}
\slugcomment{submitted to ApJ}

\shorttitle{Trigger of the X1.0 flare in the AR 12192}
\shortauthors{Bamba et al.}

\begin{document}


\title{Triggering Process of the X1.0 Three-ribbons Flare\\ in the Great Active Region NOAA 12192}


%\author{Y. Bamba, K. Kusano\altaffilmark{1}, S. Inoue\altaffilmark{2}, and D. Shiota\altaffilmark{1}}
%\affil{Institute of Space and Astronautical Science (ISAS), Japan Aerospace Exploration Agency (JAXA), 3-1-1 Yoshinodai, Chuo-ku, Sagamihara, Kanagawa 252-5210, Japan}
%\email{y-bamba@nagoya-u.jp}

\author{Yumi Bamba}
\affil{Institute of Space and Astronautical Science (ISAS)/Japan Aerospace Exploration Agency (JAXA)\\ 3-1-1 Yoshinodai, Chuo-ku, Sagamihara, Kanagawa 252-5210, Japan}
\email{y-bamba@nagoya-u.jp}

\author{Satoshi Inoue}
\affil{Institute for Space-Earth Environmental Research (ISEE)/Nagoya University\\ Furo-cho, Chikusa-ku, Nagoya, Aichi 464-8601, Japan}

\author{Kanya Kusano}
\affil{Institute for Space-Earth Environmental Research (ISEE)/Nagoya University\\ Furo-cho, Chikusa-ku, Nagoya, Aichi 464-8601, Japan}

\author{Daikou Shiota}
\affil{Institute for Space-Earth Environmental Research (ISEE)/Nagoya University\\ Furo-cho, Chikusa-ku, Nagoya, Aichi 464-8601, Japan}


%\altaffiltext{1}{Institute for Space-Earth Environmental Research, Nagoya University, Furo-cho, Chikusa-ku, Nagoya, Aichi, 464-8601, Japan}

%\altaffiltext{2}{Japan Agency for Marine-Earth Science and Technology (JAMSTEC), Kanazawa-ku, Yokohama, Kanagawa, 2360001, Japan}

%\altaffiltext{2}{Max-Planck-Institut f\"ur Sonnensystemforschung, Justus-von-Liebig-Weg 3, 37077 G\"ottingen, Germany}

%% Mark off your abstract in the ``abstract`` environment. In the manuscript

\begin{abstract} %less than 250 words
Solar magnetic field in flare productive active region (AR) is much more complicated than theoretical models that assume very simple magnetic field structure. The X1.0 flare, that occurred in AR 12192 on 2014 October 25, showed complicated three-ribbons structure. To clarify the trigger process of the flare and to evaluate the applicability of a simple theoretical model, we analyzed the data form Hinode/SOT and SDO/HMI, AIA. We investigated the spatio-temporal correlation between the magnetic field structures, especially the non-potentiality of the horizontal field, and the bright structures in the solar atmosphere. As a result, we determined that the west side of the positive polarity, that is intruding a negative polarity region, is the location where the flare was triggered. It is due to the fact that the sign of magnetic shear in that region was reversed to the major shear of the AR, and the significant brightenings were observed over the polarity inversion line (PIL) in that region before the flare onset. These features were consistent with the recently proposed flare-trigger model that suggests that small Reversed Shear (RS) magnetic disturbances can trigger solar flares. Moreover, we found that the RS field was located slightly off the flaring PIL contrary to the theoretical prediction. We discuss the possibility of an extension of the RS model based on an extra numerical simulation. Our result suggests that the RS field has certain flexibility for the displacement from highly sheared PIL and that the RS field triggers more flares than we expected.
\end{abstract}

\section{Introduction} \label{sec:intro}

The triggering mechanisms of solar flares are not yet completely elucidated.
Many previous studies have attempted to reveal the physics of solar flare occurrence using theoretical and/or observational methods \citep[e.g.][]{antiochos99, chen_shibata00, moore01}.
Currently, the CSHKP model \citep{carmichael64, srurrock66, hirayama74, kopp_pneuman76} is widely accepted as a standard solar flare model.
In this model, magnetic reconnection plays an important role in the energy release process of the solar flare, and it is substantiated by the Yohkoh satellite \citep{ogawara92} observations.
Yohkoh captured a bright cusp structure in the solar corona with soft X-ray observations \citep{tsuneta92}, and it is believed to be evidence of energy release by magnetic reconnection \citep{tsuneta96}.
However, we still do not know what triggers the magnetic reconnection process although there are several candidates such as emerging flux \citep[e.g.][]{heyvaerts77, kurokawa02}, flux cancellation \citep[e.g.][]{zhang01, green11}, sunspot rotation \citep[e.g.][]{louis14}, shear motion \citep[e.g.][]{hagyard84, kusano95}, and the reversal of shear \citep{kusano04}.
The quantitative conditions of each flare trigger model have not yet been clarified.
It might be cause our limited capacity for flare forecast.

\citet{kusano12} performed a new numerical simulation to investigate the trigger of solar flares.
They used a simple magnetic field structure, and characterized it using two parameters: the global magnetic shear angle $\theta$ of the active region (AR) and the azimuth $\phi$ of the small bipole appearing on the polarity inversion line (PIL, see Figure 1 in \citet{kusano12} and \citet{bamba13}).
They surveyed which combination of $\theta$ and $\phi$ can trigger flares, and found that two types of structures, the Opposite Polarity (OP) and the Reversed Shear (RS) type, are capable of triggering flares.
\citet{bamba13} performed an observational verification of the flare trigger model proposed by \citet{kusano12}.
They analyzed four major flares observed by the Hinode satellite \citep{kosugi07}, and quantitatively confirmed how the observed results of $\theta$ and $\phi$ are consistent with Kusano et al.'s numerical simulation results.
They identified small magnetic bipoles triggering flares in ARs with relatively simple structure of magnetic field and the PIL.
However, solar flares generally occur in complicated ARs that are very different from the simple structure assumed in the simulation. 
Therefore, it is important to examine whether the theoretical model is applicable to more complex ARs.

AR NOAA 12192 was the biggest sunspot region in the solar cycle 24.
The AR stayed on the solar disk through three Carrington rotations, and it was numbered NOAA 12172 (and 12173), 12192, and 12209 in each phase.
AR 12192 had the most complicated magnetic structure including a $\delta$-type sunspot in which several umbra are sharing penumbra.
A major bipole extended from east to west, and emerging flux was seen in the middle of the bipole.
Six X-class flares occurred in the AR during disk passage, and these showed clear flare ribbons appearing in the central region of the AR.
Especially, the X1.0 flare on 2014 October 25, which is the fourth X-class flare in the AR, produced three flare ribbons.

\citet{kusano12} and \citet{bamba13} have demonstrated that the small magnetic bipoles that trigger flares are located in the center of the initial flare ribbons.
However, it is difficult to infer where in the AR the triggering structure is and to reveal the physical processes for the three-ribbons flare.
In this paper, we aim to clarify the X1.0 flare trigger process in AR 12192 to examine whether the trigger model is applicable even to complicated ARs. 
Moreover, it should be noted that neither the X1.0 flare nor the other X-class flares in the AR produced any coronal mass ejections (CMEs) although it is statistically reported that 75\% of X-class flares are associated with CMEs \citep{yashiro06}.
We also discuss why the X1.0 flare did not have associated CMEs from the point of view of the triggering mechanism.

This paper is organized as follows.
The sampled data and analysis method are described in Sections~\ref{sec:data} and \ref{sec:method}, respectively.
The data analysis results are shown in Section~\ref{sec:result}.
The conceivable scenarios of the X-class flare and the reason why the flares in the AR did not associated with any CMEs are discussed in Section~\ref{sec:discussion}.
Finally, we summarize the results and considerations in Section~\ref{sec:summary}.

\section{Data descriptions} \label{sec:data}

\begin{figure*}
\epsscale{2.00}
\plotone{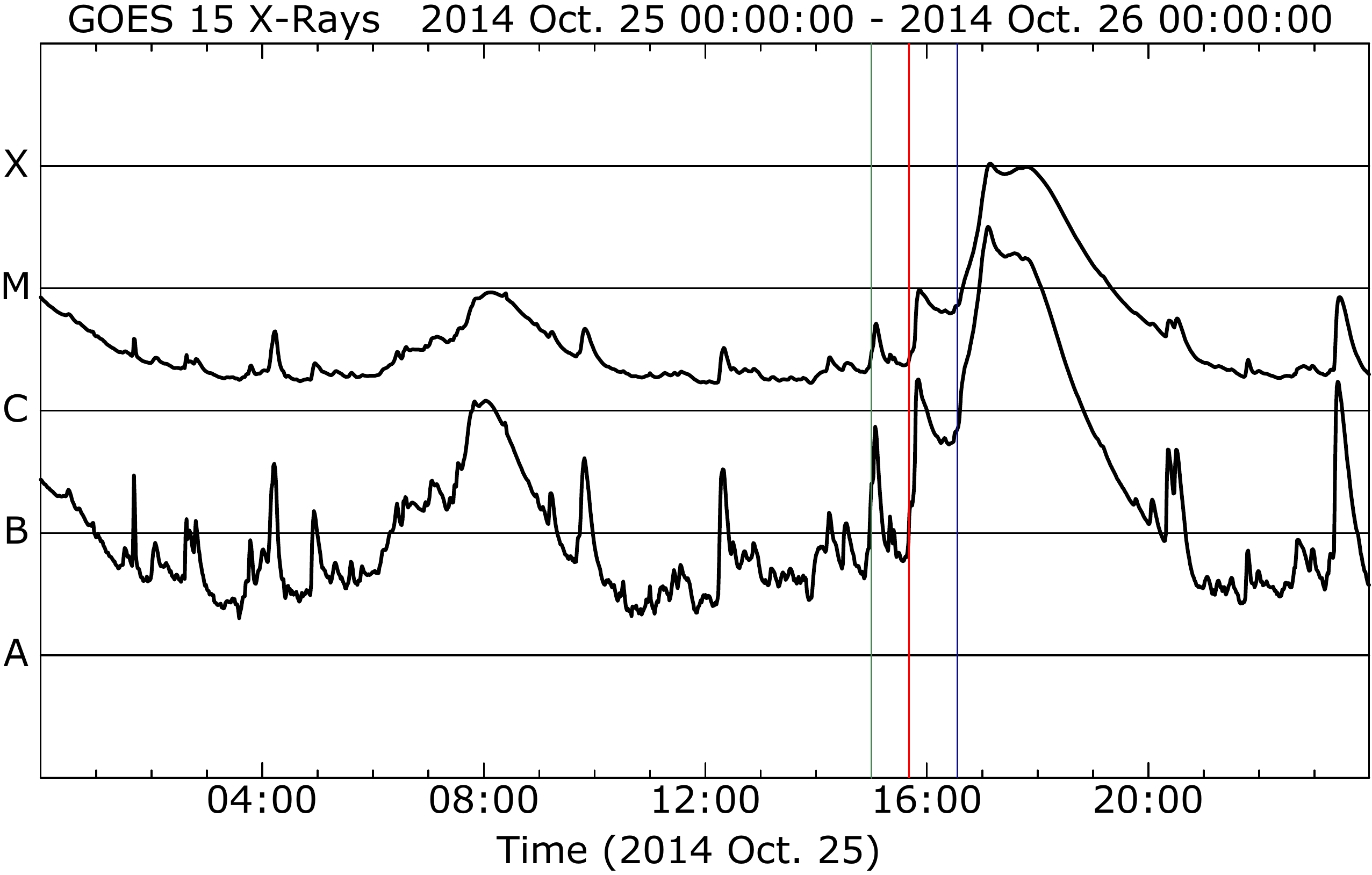}
\caption{
The soft X-ray light curve observed by GOES (1-8 {\AA} and 0.5-4 {\AA}) from 00:00 UT on 25 October to 00:00 UT 26 October 2014.
The green/red/blue vertical line indicates the onset time of the C5.1 (15:00 UT)/C9.7 (15:44 UT)/X1.0 (16:31 UT) flare, respectively.
}
\label{fig:goes_curve}
\end{figure*}

More than 130 flares, which were larger than C1.0, occurred in the great AR 12192.
Table~\ref{table:eventlist} lists the X-class flares that occurred in the AR.
Solar Dynamics Observatory (SDO; \citet{pesnell12}) observed all of the flares whereas Hinode/SOT observed latter five.
In this paper, we focused on the X1.0 flare on 2014 October 25, which was observed by both Hinode and SDO.
The AR was located at S$10^{\circ}$-$20^{\circ}$ latitude and W$15^{\circ}$-$25^{\circ}$ longitude during the flare.
Figure~\ref{fig:goes_curve} shows the soft X-ray light curve observed by GOES (1-8 {\AA} and 0.5-4 {\AA}).
The onset time of the X1.0 flare was 16:31 UT \footnote{This time is defined from X-ray observation by the RHESSI satellite \citep[Reuven Ramaty High Energy Solar Spectroscopic Imager;][]{lin02} because GOES likely released wrong times for this event.}, and it is indicated by a blue vertical line in Figure~\ref{fig:goes_curve}.
The two C-class flares (C5.1 at 15:00 UT and C9.7 at 15:44 UT) occurred before the X1.0 flare onset, and these are indicated by the green and red vertical lines.

We analyzed SDO data from 13:00 UT to 19:00 UT on October 25.
SDO observes the full disk of the Sun (2000$^{\prime\prime}$ $\times$ 2000$^{\prime\prime}$), and we used filter magnetograms taken by Helioseismic and Magnetic Imager (HMI; \citet{schou12}) at 6173 {\AA} (\ion{Fe}{1} line, sample the photospheric magnetic field), filtergrams taken by Atmospheric Imaging Assembly (AIA; \citet{lemen12}) at 1600 {\AA} (continuum and \ion{C}{4} line, sample upper photosphere and the transition region), 171 {\AA} (\ion{Fe}{9} line, sample upper transition region and quiet corona), 131 {\AA} (\ion{Fe}{8} and \ion{Fe}{21} lines, sample the transition region and flaring corona), 193 {\AA} (\ion{Fe}{12} and \ion{Fe}{24} lines, sample the corona and hot flare plasma), and 304 {\AA} (\ion{He}{2} line, sample the chromosphere and the transition region).
%The region of the atmosphere and related formation temperatures of each wavelength are summarized in Table~\ref{table:aia_lines} \citep[excerpt from ][]{lemen12}.
The spatial resolution and the cadence were 1$^{\prime\prime}$ and 45 seconds for HMI, 1.5$^{\prime\prime}$ and 12 seconds for AIA (except AIA 1600 {\AA} where the cadence was 24 seconds).
We investigated the spatial and temporal correlation between the evolution of the photospheric magnetic field, brightenings in the chromosphere, the transition region, and the corona.

Moreover, we used vector magnetic field data (the Spaceweather HMI Active Region Patch:SHARP) obtained by SDO/HMI on October 25 15:00 UT to calculate the NLFFF.
We also used Hinode/SOT data to perform more precise analysis of the magnetic field.
We used the full polarization states (Stokes-I, Q, U, and V) at 6301.5 and 6302.5 {\AA} (\ion{Fe}{1} line) with a sampling of 21.5 {m\AA}, obtained by the Spectro-Polarimeter (SP).
The SP scanned the central part of the AR at 11:00-11:33 UT with a 164$^{\prime\prime}$ $\times$ 164$^{\prime\prime}$ field-of-view (FOV), and the spatial resolution was 0.3$^{\prime\prime}$.

\section{Analysis Methods} \label{sec:method}

\subsection{Method of SDO/HMI, AIA analysis} \label{sec:method_sdo}

In this study, we basically used the analysis method developed in \citet{bamba13}.
The analysis method was originally developed for Hinode/SOT data, but \citet{bamba14} already examined the applicability of the analysis method to the SDO data sets.
Here, we briefly summarize the procedures of the analysis method.

We used HMI level 1.5 line-of-sight (LOS) magnetograms ({\tt hmi.M\_45s} series) and AIA level 1.0 data ({\tt aia.lev1\_euv\_12s} and {\tt aia.lev1\_uv\_24s} series).
We first calibrated all the HMI filter magnetograms and all the AIA images using the {\tt aia\_prep} procedure in the Solar Soft-Ware (SSW) package.
By this process, spatial fluctuations were reduced, and the images were rotated so that the solar EW and NS axes are aligned with the horizontal and vertical axes of the image, respectively.
Moreover, LOS magnetograms and AIA images were resampled to the same size because the pixel scales are different between HMI and AIA.
Thus the positions of the LOS magnetograms and AIA images were aligned.
Next, we chose a HMI filter magnetogram and an AIA image closest in time, and these two images were superimposed onto each other.
We drew the PILs and the strong emission contours in the AIA image onto the LOS magnetogram at each time.
The PILs were drawn on the AIA images at each time, as well.

\subsection{Method of Hinode/SOT analysis} \label{sec:method_hinode}

The SP scan data were calibrated using the {\tt sp\_prep} procedure \citep{lites_ichimoto13} in the SSW package assuming the Milne-Eddington atmosphere.
The inversion code MEKSY (developed by Dr. Takaaki Yokoyama) was adopted and the $180^{\circ}$ ambiguity in the vector magnetograms is resolved using the AZAM utility \citep{lites95}.
Note that the observed ``LOS'' and ``transverse'' magnetic field vectors (i.e., image plane magnetic field) converted to the heliographic magnetic field $B_{x}$, $B_{y}$, and $B_{z}$ using the equation (1) of \citet{gary_hagyard_1990}.

We investigated the distribution of the magnetic shear over the AR before the flare onset, using the vector magnetogram obtained from the SP scan data.
We first calculated the potential field using the {\tt fff} procedure in the {\tt nlfff} package (developed by Dr. Yuhong Fan) in SSW.
Then we measured the angles between the potential field vector ${\bm B_{p}}$ and horizontal field vector $\bm B_{h} = \sqrt{{B_{x}}^2 + {B_{y}}^2}$ in each pixel, and we defined these angles as the relative shear angle $\chi$.
The relative shear angles are defined between $\pm180^{\circ}$, where $0^{\circ}$ means vectors ${\bm B_{p}}$ and ${\bm B_{h}}$ are oriented in the same direction.
The direction of vector ${\bm B_{h}}$ deviates from vector ${\bm B_{p}}$ to counter-clockwise (clockwise) when the magnetic helicity is positive (negative) and the value of $\chi$ is positive (negative).
We colorized the relative shear angles as shown in Section~\ref{sec:result_SP}.

\section{Results} \label{sec:result}

\subsection{Overview of the C5.1, C9.7 and X1.0 flares in each wavelength} \label{sec:result_overview}

The temporal evolution of the brightenings and flare ribbons in AIA 1600 {\AA}, which is sensitive to the emission from upper chromosphere and the transition region, are shown in Figure~\ref{fig:AIA1600}.
We can see the contours of those brightenings with the magnetic field structure in Figure~\ref{fig:HMI1600}, where white/black indicates the positive/negative polarity of the LOS magnetic field.
The green lines and the red contours denote the PIL (line of 0 G) and the significant brightenings (2000 DN) in AIA 1600 {\AA}, respectively.
The green lines in Figure~\ref{fig:AIA1600} are the same as these in Figure~\ref{fig:HMI1600}.
The positive polarity that is intruding the negative region N1 (henceforth IPP: intruding positive polarity) and a weak negative region N2 are located between the major bipole P1 and N1, as can be seen in Figure~\ref{fig:HMI1600}(a).
We can also see the small brightenings B1 and B2, which are indicated by the yellow arrows in panels (a) of both Figures~\ref{fig:AIA1600} and \ref{fig:HMI1600}.
These brightenings are seen intermittently at the west side of the IPP at 14:48-14:58 UT.
The other brightening B3 is also seen in the IPP in panels (a).
The flare ribbons of the C5.1 flare appears about five minutes later as shown in panels (b).
The positive ribbons CR1 and CR2 are clearly seen in the P1 and IPP regions.
There are two negative ribbons in the N1 region at a glance, but these brightenings are connected each other in the AIA 304 {\AA} image (see Figure~\ref{fig:AIA304}(b) and latter description).
Then we identify these brightenings in the N1 region as the negative ribbon CR3 as seen in Figure~\ref{fig:AIA1600}(b).
The B1 and B2 disappear at the last minute before the C5.1, while the B3 enhances as the flare ribbon CR2 in panels (b).
It is inferred that the B1 and B2 are caused by local magnetic reconnection occurred at the west side of the IPP before the C5.1 flare onset, and that the B1 and B2 brightenings appear at the foot points of the small magnetic loops connecting the IPP and N2.
It is also suggested that the B3 locates at a foot point of magnetic loops that connects the IPP and N1 and that corresponds to the C5.1 flare.
Once the three ribbons CR1, CR2, and CR3 of the C5.1 flare disappear by 15:15 UT, the C9.7 flare ribbons appear from 15:40 UT as seen in panel (c).
The C9.7 flare also shows three flare ribbons CR1, CR2, and CR3 at the same location to the ribbons of the previous C5.1 flare.
The three-ribbons of the C5.1 and C9.7 flares almost never propagate with time.
A faint ribbon-like brightening remained at the region where the CR1 is seen even after the three ribbons disappeared, as seen in  Figure~\ref{fig:AIA1600}(d).
Panels (e) show the initial flare ribbons of the X1.0 flare that appear about one hour after the C9.7 flare.
In this phase, the XR1 (positive ribbon) and the XR2 (negative ribbon) at the west side/middle are more clearly seen than the XR3 (negative ribbon) at the east side.
These three-ribbons slowly grow as shown in panels (f) at 16:35-17:10 UT.
Especially, XR1 and XR3 get longer in the southward and northward directions, respectively.
However, they almost never propagate to outer side (i.e. to the westward and eastward) with time.
The XR2 disappears first in the late phase while XR1 and XR3 remain more than one hour after the onset.

Figure~\ref{fig:AIA304} shows the AIA 304 {\AA} images of the same FOV and almost the same time as Figures~\ref{fig:AIA1600} and \ref{fig:HMI1600}.
The PILs are over plotted only in panel (a).
The intermittent brightenings B1, B2, and B3 are indicated by the yellow arrows in panel (a), and these are seen at the west side of the IPP as well as shown in the AIA 1600 {\AA} image of Figure~\ref{fig:AIA1600}(a).
The brightening B1 and B2 are connected each other and it shapes a small loop striding over the local PIL at the west side of the IPP.
It is clearly seen at 14:48-15:00 UT although it persisted from 13:00 UT to just before the onset of the C5.1 flare.
It also suggests the existence of small magnetic loops which connect the IPP and N2 in the west side of the IPP.
On the other hand, the brightening B3 is connected with the N1 where the CR3 will appear, as clearly seen in Figure~\ref{fig:AIA304}(a) and (b).
The brightenings, these will be the flare ribbons CR1 and CR2, are already seen from 13:00 UT, and the CR3 also appear from 14:00 UT.
The three ribbons start to be enhanced from 14:30 UT, and these will be the three ribbons of the C5.1 flare in panel (b).
There is a filament on the PIL between P1 and N2, and some bright points start to move along the filament from 
just before the C5.1 flare (from 14:58 UT).
The three ribbons CR1, CR2, and CR3 remain even after C5.1 flare, and the bright point motion along the filament continues and become intense.
These bright structures flow down along the filament to the foot points in the P1 and N1 regions, and then the three ribbons CR1, CR2, and CR3 of the C9.7 flare are enhanced from 15:47 UT (see panel (c)).
Note that the strong brightening, that looks like another flare ribbon between the CR1 and CR2 in panel (c), is not a flare ribbon, but a bright point moving along the filament like the tether-cutting reconnection (cf. Figure~\ref{fig:AIA171} and latter description).
The flare ribbons remain more clearly than that seen in Figure~\ref{fig:AIA1600}(c), especially the CR1 and CR3 are clearly seen in Figure~\ref{fig:AIA304}(d).
The remaining ribbons are enhanced and a new flare ribbon along the filament gradually appeared in the N2 region from 16:30 UT, and it becomes XR2 as seen in panel (e).
This is the X1.0 flare, and the flare ribbons XR1 and XR3 are corresponding to the CR1 and CR3, respectively.
The CR2 also remain, but it is not seen in the AIA 1600 {\AA} images (see Figure~\ref{fig:AIA1600}(e, f)).
In the initial phase of the X1.0 flare (until 16:55 UT), ex-CR2 brightening and negative ribbon XR3 are connected, as seen in Figure~\ref{fig:AIA304}(e).
It is corresponding to the first peak of the X1.0 flare in the GOES soft X-ray light curve (cf. Figure~\ref{fig:goes_curve}).
Then the bright bridge connecting ex-CR2 and XR3 become weakened, and the XR3 get longer to the northward as seen in panel (f).
The XR1 and XR2 are also enhanced at that time, and the post flare loop connecting XR1 and XR2 is appeared.
The XR2 is weakened first while the XR1 and XR3 remain likewise seen in AIA 1600 {\AA} images.

The coronal loops in AIA 131 {\AA} images are shown in Figure~\ref{fig:AIA131}, with the PILs only shown in panel (a).
In the early phase (panel (a)), the bright loops L1 and L2 connect IPP-N1 and P1-N1, respectively.
The IPP have continuously emerged from October 19, and it has intruded into the negative sunspot (N1).
Therefore, it is suggested that strong electric current layer is formed in between the L1/L2 and the overlaying magnetic arcade which connects P1-N1 (it locates higher/outer than L2).
In this layer, coronal loops are heated by magnetic reconnection between L1/L2 and overlaying magnetic arcade, even though the field directions of magnetic loops do not differ strongly.
Note that there is another faint loop L3 as indicated in panel(a), and it might connects P1-N2 (and also P1-N1).
The foot points of L1 and L2 brighten as seen in Figure~\ref{fig:AIA304}(a).
These heated loops L1 and L2 are enhanced at the C5.1 flare onset as seen in Figure~\ref{fig:AIA131}(b).
Once the intensity of the coronal loops decrease at 15:15-15:35 UT, the loops L1 and L2 becosme bright again and the three ribbons of the C9.7 flare appear as indicated by the yellow broken lines in panel (c).
Then the intensity of the coronal loops continuously increase until the onset time of the X1.0 flare (panel (d)).
The coronal loops connecting the XR1 and XR2 enhance (panel (e)), then the XR3 appear and the faint coronal loops connect the XR1 and XR3 as can be seen in panel (f).

Interestingly, the tether-cutting magnetic reconnection process \citep{moore01} is clearly seen between the C5.1 and C9.7 flares.
Figure~\ref{fig:AIA171} shows AIA 171 {\AA} images with a similar FOV to Figures~\ref{fig:AIA1600}-\ref{fig:AIA131}.
The PILs are over plotted only in panel (a), and there is a filament as indicated by the red arrow in panel (b).
We can see the faint loop structures which intersect at point O and along the filament.
Some small brightenings frequently move along these faint loops as illustrated by the blue arrows in panel (b).
These bright structures are seen from 13:00 UT, and these are enhanced from 15:00 UT, just before the C5.1 flare.
The motion along the loops become faster and the loops slightly expand between 15:40 and 16:00 UT (panel (c)).
The flare ribbons CR1 and CR3 of the C9.7 flare appear at the foot points of the faint loops as illustrated by red broken lines in panel (d), and we can see the three ribbons in the chromosphere at that time (see Figure~\ref{fig:AIA304}(b)).
These are clearly seen in Movie 1.
The similar motion of the bright structure along the filament is also seen in AIA 304 {\AA} images from just after the C5.1 flare to the onset of the C9.7 flare.
It suggests that the tether-cutting magnetic reconnection occurs in the C5.1 and C9.7 flares at the point O.

The bright points B1, B2, B3, the tether-cutting reconnection, and the bright coronal loops connecting the XR1 and XR2 are also seen in AIA 193 {\AA} images (Figure~\ref{fig:AIA193}) with much less saturation.
In panel (a), the brightenings B1, B2, and B3 are clearly seen as well as seen in AIA 1600 {\AA} and 304 {\AA} images.
The brightening B3 corresponds to the positive foot point to the coronal loop L1 that connect the IPP and N1 (cf. Figure~\ref{fig:AIA131}(a)).
The positive foot point of L2 is also seen as a bright point in P1 region, as seen in Figure~\ref{fig:AIA193}(a).
The tether-cutting reconnection is also seen as panel (b), although it is more faint than that seen in AIA 304 {\AA} and 171 {\AA} images.
After the tether-cutting reconnection, from 16:30 UT, the faint coronal loops rooted the XR1 and XR2 drastically expand to the southwest ward, as illustrated by the blue arrow in panel (c).
This expanding motion of the faint coronal loops correspond to the enhancement of the coronal loops seen in Figure~\ref{fig:AIA131}(d-f). 
Then the flare ribbons XR1 and XR3 are elongated to the southward and the northward, respectively.

\subsection{Locations of the C9.7 and the X1.0 flare ribbons} \label{sec:result_ribbons}

The flare ribbons of the two C-class flares (the C5.1 and C9.7 flares) appear at the same location, as we described in the previous section.
Therefore, here we compare the locations and evolution between the C9.7 and the X1.0 flare ribbons.
Figure~\ref{fig:flare_ribbons} shows the distribution of the C9.7 and X1.0 flare ribbons.
The background images are the HMI LOS magnetograms at the onset time of the X1.0 flare (16:31 UT), and the green lines indicate the PILs.
The red contour outlines the brightenings in AIA 1600 {\AA}, such as the flare ribbon at 15:50 UT (just after the C9.7 flare onset), in both panels (a) and (b).
The blue contours outline the brightenings at 16:45 UT (before the X1.0 flare onset) in panel (a) and at 17:06 UT (after the X1.0 flare onset) in panel (b), respectively.
The initial flare ribbons of the C9.7 and X1.0 flare are seen in panel (a), and the XR1 and XR3 of the X1.0 flare located slightly to the outer side of the CR1 and CR3 of the C9.7 flare.
Both of the XR1, XR3, CR1, and CR3 slightly propagate to the outer side (i.e. to westward and eastward) with time, although the XR1 and XR3 clearly grow to southward and northward, as noted in Section~\ref{sec:result_overview}.
Therefore, it is suggested that the C5.1, C9.7, and X1.0 flares serially occur corresponding to almost the same magnetic structures.
In other words, the X1.0 flare is an extension of the C-class flares, and the X1.0 flare is driven by reconnection of the magnetic field anchored in the outer region of the C-class flare ribbons.

\subsection{Features of the magnetic fields} \label{sec:result_SP}

The vector magnetic field in the central part of the AR was observed by Hinode/SP between 11:00 and 11:33 UT, as shown in Figure~\ref{fig:SP}.
In panel (a), the background white/black indicates positive/negative polarity of $B_{z}$.
Green lines indicate the PILs (line of $B_{z} = 0 G$).
The horizontal magnetic field vector $\bm B_{h}$ is over plotted on the $B_{z}$ map by the red arrows.
The horizontal field is strongly sheared along the PIL located between the P1 and N2 regions.
It suggests that the major magnetic helicity along the PIL of the AR is negative.
On the other hand, the vectors at the west side of the IPP are locally toward northwest stride over the PIL.
It is consistent with the shape of the small bright loop connecting B1 and B2 in Figure~\ref{fig:AIA304}(a).
Therefore, the local magnetic field at the west side of the IPP has positive magnetic helicity, which is opposite to the major magnetic helicity along the PIL.
It is more clearly seen in panel (b), which shows the relative shear angle $\chi$, defined as the angle between the potential field $\bm B_{p}$ and the horizontal field $\bm B_{h}$ at each point.
The black lines indicate the PILs, that is the same to the green lines in panel (a).
The relative shear angle $\chi$ along the flaring PIL (between P1 and N2) and that in the west side of the IPP are correspond to the shear angle $\theta_{0}$ of the AR and the azimuth $\varphi_{e}$ of the small magnetic disturbance in \citet{kusano12}, respectively.
Obviously, $\chi$ is around $-90^{\circ}$ (blue) along the PIL but $\chi$ is $90^{\circ}$ (red) at the west side of the IPP.
Therefore, the distribution of the relative shear angle $\chi$ suggests that the west side of the IPP satisfied the Reversed Shear (RS) type flare trigger condition proposed by \citet{kusano12}.
They proposed that the shear cancellation between the global magnetic field and the RS-type small magnetic field structure can trigger a flare, and significant brightenings should be observed in the solar atmosphere during the shear cancellation over the RS-type structure.
In our case, significant brightenings B1 and B2 are observed at the west side of the IPP where the magnetic shear is reversed to the shear of the global magnetic field along the flaring PIL.
It is consistent with the theoretical prediction.
Therefore, the west side of the IPP could be the location where the flares were triggered in the AR.

\section{Discussion} \label{sec:discussion}

%\subsection{Flare trigger scenario: comparison with the simulation models} \label{sec:discussion_scenario}
\subsection{Coronal magnetic field extrapolation} \label{sec:discussion_nlfff}

In this study, we extrapolated the magnetic field in the corona, and compared it to the observed features.
We derived the coronal magnetic field from vector magnetic field data taken by HMI at October 25 15:00 UT using the NLFFF extrapolation method developed by \citet{inoue14}.
Figure~\ref{fig:nlfff} shows the coronal magnetic field lines anchor to the flare ribbons of the C9.7 and the X1.0 flares.
The grayscale is an HMI LOS magnetogram at October 25 15:00 UT.
The red/blue contour outlines the brightenings (700 DN) in AIA 1600 {\AA}, such as the flare ribbons observed at 15:50 UT/17:03 UT.
The blue and orange tubes indicate the coronal magnetic field lines.
The orange tubes anchor to the ribbons of the C9.7 flare (CR1, CR2, and CR3), as shown in panels (a) and (b).
On the other hand, the sky blue tubes anchor to the three ribbons of the X1.0 flare (XR1, XR2, and XR3), as seen in panels (c) and (d).
The small magenta tubes indicated by the magenta arrow are the local magnetic field lines at the west side of the IPP.
Obviously, the sky blue magnetic loops connecting XR1 and XR3 are located at the outer edge of the orange loops connecting CR1 and CR3.
These are consistent with the observed features that the flare ribbons of the X-class flare are on the outer side of the ribbons of the C-class flares as shown in Figure~\ref{fig:flare_ribbons}.

\subsection{Comparison with the emerging flux model} \label{sec:discussion_emerging}

Here, we discuss the flare trigger scenario of the consecutive C- and X-class flares by comparing a theoretical model proposed by \citet{chen_shibata00}.
In this region, there are four kinds of possible magnetic connectivity; P1-N1, IPP-N1, IPP-N2, and P1-N2.
Moreover, there is a filament trapped by P1-N2 loops as seen in AIA images (Figures~\ref{fig:AIA304}, \ref{fig:AIA171}, and \ref{fig:AIA193}).
If we focus on the connectivity IPP-N1 and P1-N2, the topology of the magnetic field is consistent with the case B of \citet{chen_shibata00}, in which the emerging flux appears on the outer edge of the filament channel.
The IPP-N1 and P1-N2 has the same orientation relative to the main flaring PIL each other, and the IPP-N1 can be considered as an  flux emerging on the east edge of the filament channel P1-N2.
Magnetic reconnection of the IPP-N1 flux with the flux passing over the main flaring PIL (between P1 and N2) will weaken the latter flux.
It can destabilize the current-carrying flux in the core region above the main flaring PIL, and the reconnection also produces short loops at the PIL between the IPP-N2 and long loops connect P1-N1.

In this scenario, it is required to trigger an X-class flare that the flux rope exists on the PIL and that the flux rope erupts.
The former is likely because multitude of flares including three X-class flares had occurred in the same PIL between P1-N2.
%However, the latter is not consistent with the result of analysis based on the NLFF field.
%Figure~\ref{fig:DI1p5}(a) is the side view of Figure~\ref{fig:nlfff}(f) in which the yellow colored surface is an isosurface where the decay index equals to 1.5.
%The decay index is defined as $n = -{d\ln{\mid{\bm B}\mid/}d\ln{\mid{z}\mid}}$, where ${\bm B}$ and $z$ are the potential magnetic field ${\bm B_{p}} = (B_{px}, B_{py}, B_{pz})$ (calculated from the observed $B_{z}$) and the height, respectively \citep[e.g.][]{shafranov66, kliem_torok06}.
%Obviously, the highly sheared loops corresponding to the flares in the AR do not reach the isosurface.
However, the latter is not consistent with the observed features that the filament along the PIL between P1-N2 does not move through the two C-class and X-class flare processes (cf. Figure~\ref{fig:AIA304}).
It indicates that the eruption of flux rope such as required in the case B scenario is unlikely.
Moreover, IPP is intruding into N1, i.e magnetic loop connecting IPP-N1 moves away from P1-N2, in association with emergence of IPP, as described in Section 4.1. This motion is oppose to the scenario expected in the case B of \citet{chen_shibata00} that requires approach of IPP-N1 and P1-N2.
Therefore, the flare trigger by the case B scenario may be difficult, although the magnetic field topology is consistent with the case.

\subsection{Comparison with the RS-type model} \label{sec:discussion_RS}

As an another flare trigger scenario, there is the RS-type scenario proposed by \citet{kusano12}.
In this senario, the key structure is the small magenta loops connecting IPP-N2, that are indicated by the magenta arrow in Figure~\ref{fig:nlfff}.
We found that the west side of the IPP satisfied the condition of the RS-type flare trigger field based on the distribution of the relative shear angle (cf. Figure~\ref{fig:SP}(b)).
%There are small magenta loops in the west side of the IPP as indicated by the magenta arrow in Figure~\ref{fig:nlfff}, and these can be a RS-type flare trigger field.
%The magnetic field topology in Figure~\ref{fig:nlfff}(e) is the same as that in Figure 4(b, c) of \citet{kusano12}.
The long orange loops connecting P1-N1 are topologically equivalent to the overlying blue field line in Figure 4(a-c) of \citet{kusano12}.
Therefore, the following scenario is conceivable.

%first reconnection (trigger for the C-class flares)
(1) Step-1 (trigger for the C-class flares):
Magnetic shear cancellation proceeds between the IPP-N2 (magenta) loops and the P1-N1 (orange) loops at the west side of the IPP.
The shear canceling reconnection locally heats the atmosphere and causes chromospheric evaporation, above the west side of the IPP.
These local heating and evaporation are observed as the brightening B1, B2, and B3 in AIA images (Figures~\ref{fig:AIA1600}-\ref{fig:AIA304}).
The reconnected fluxes are transferred into the small loops connecting IPP-N1 and P1-N2, and the magnetic pressure above the west side of the IPP must be decreased.
The loops IPP-N1, P1-N1, and P1-N2 are heated by the reconnection.
These are observed as bright coronal loops L1, L2, and L3 in AIA 131 {\AA} (Figure~\ref{fig:AIA131}), although the location of the negative foot point of L3 is fuzzy.
The schematics of the magnetic field line before and after the step-1 reconnection are shown in Figure~\ref{fig:schematic}(a, b).
The orange and magenta loops are equivalent to the P1-N1 (orange loops) and IPP-N2 (magenta loops) loops in Figure~\ref{fig:nlfff}(a, b).
The green small loops in Figure~\ref{fig:schematic}(b) represent the small loops formed by the magnetic reconnection between the RS-type trigger field (magenta loops) and the overlying sheared field (orange loops).

%second reconnection (C-class flares)
(2) Step-2 (reconnection of the C-class flares):
The sheared loop P1-N1 collapses inward because an inflow (such as studied by \citet{ugai_shimizu96}) drags flux into the region where magnetic pressure was decreased by the step-1 reconnection, i.e., above the west side of the IPP.
Then reconnection occurs in the collapsed loops (inside the orange loops connecting P1-N1 in Figures~\ref{fig:nlfff}, \ref{fig:schematic}(b)), and it shows the three-ribbons CR1, CR2, and CR3 of the C5.1 and C9.7 flares.
%Note that only one ribbon appears in the positive polarity region (P1) while two ribbons appear in the negative polarity region (N1) because both the loops P1-N1 and P1-N2 root the P1 region.
The reconnection is observed as the tether-cutting reconnection in AIA 171 {\AA} (Figure~\ref{fig:AIA171} and Movie 1) and 193 {\AA} (Figure~\ref{fig:AIA193}) images.
Note that the CR2 in the IPP was the brightening B3 in the pre-flare phase, and that the brightening in the IPP represents the intensive electric current layer formed by the step-1 reconnection (see Appendix).

%third reconnection (trigger for the X-class)
(3) Step-3 (trigger for the X1.0 flare):
It is inferred that the long twisted flux rope connecting P1-N1 might be formed by the tether-cutting reconnection, as illustrated by the thick orange line in Figure~\ref{fig:schematic}(c), according to \citet{moore01} and \citet{kusano12}.
There are another sheared loops (sky blue loops connecting P1-N1 and P1-N2) above the orange loops and the long twisted flux rope.
The flux rope gradually rise upwards with the overlying sky blue loops, then the magnetic fluxes are transferred into upward and magnetic pressure over the PIL between the P1 and N1 regions is decreased.
After the magnetic pressure is decreased sufficiently, magnetic reconnection between the sky blue loops occurs.
It causes the X1.0 flare and produces the three-ribbons (XR1, XR2, and XR3) at the foot points of the sky blue loops.
The reconnection might produce further long twisted flux rope connecting P1-N1, then the flare ribbons XR1 and XR3 are elongated to the south and north as magnetic reconnection with different field lines.
%Therefore, the C5.1, C9.7, and X1.0 flares can be triggered originally by the RS-type structure (the magenta loops connecting IPP-N2) at the west side of the IPP in a three-step process.

Therefore, the C5.1 and C9.7 flares can be triggered originally by the RS-type structure (the magenta loops connecting the IPP-N2) at the west side of the IPP, and the rise of the flux rope that formed by the C-class flares trigger the X1.0 flare, in a three-step process.
In \citet{kusano12}, they put a small bipole, that satisfy the RS-type condition, just above the highly sheared PIL.
Then the shear cancellation (step-1 reconnection) between the trigger field and the pre-existing sheared field occurs on the PIL.
%It is expected by the original simulation of the KB12 model that the flare ribbons appear as sheared two-ribbon on both sides of the PIL, and that the trigger region locates at the center of the sheared ribbons.
%\citet{bamba13} found several flare trigger fields located at the center of the sheared flare ribbons.
%These studies treated magnetically simple ARs.
However, in the current AR, the IPP-N2 magenta loops are not located on the PIL between the major flare ribbons XR1 and XR2, i.e., the flare trigger field is located slightly off the main flaring PIL.
%The shear cancellation and the flare reconnection occurs at different PILs, and, correspondingly, the flare reconnection is topologically different from the original RS model.
%Therefore, a new question is rose up: Does the IPP-N2 magenta arcade can work as a RS-type flare trigger?
In the Appendix, we discuss the applicability of the RS-type scenario in the case of that the trigger field exist away from the main flaring PIL.
According to an extra simulation, RS-type field can work as a trigger of the flare even it is located away from the highly sheared PIL.
It is suggested that the RS-type flare trigger process has certain flexibility for the distance of the flare trigger field from the highly sheared PIL.
%Therefore, here we discuss whether the observed features are consistent with the RS-type flare trigger process even the trigger field is not located on the highly sheared PIL.

\subsection{Interpretation of ``No CME occurrence''} \label{sec:discussion_CME}

Figure~\ref{fig:DI1p5}(a) is the side view of Figure~\ref{fig:nlfff}(f) in which the yellow colored surface is an isosurface where the decay index equals to 1.5.
The decay index is defined as $n = -{d\ln{\mid{\bm B_{p, h}}\mid/}d\ln{\mid{z}\mid}}$, where ${\bm B_{p, h}}$ and $z$ are the horizontal component of the potential magnetic field calculated from the observed $B_{z}$ and the height, respectively \citep[e.g.][]{shafranov66, kliem_torok06}.
%Obviously, the highly sheared loops corresponding to the flares in the AR do not reach the isosurface.
%The yellow surface in Figure~\ref{fig:DI1p5}(a) represents the isosurface where the decay index equals to 1.5, and it corresponds to the critical height of the torus instability.
Obviously, the coronal magnetic loops anchored to both the C9.7 and the X1.0 flare ribbons have not reached the isosurface of {\it n} = 1.5, and the flux tube was stably staying because the overlying magnetic field above the flux tube did not  reach the critical decay index.
Therefore, the highly sheared magnetic loops, that are related to the flares, have not satisfied a condition of the torus instability although they satisfy the flare trigger condition  summarized in \citet{inoue15}.
In other words, the closed loops (red loops in panel (b)), that are over the orange and the sky blue loops, might prohibit the twisted flux rope, that is created in the central part of the AR via tether-cutting reconnection with the C- and X-class flares, from erupting \citep{Inoue16}.

According to \citet{kusano12}, the causality between the flux rope eruption, which may create CMEs, and the magnetic reconnection of flares is different between the two types of flare trigger processes.
The RS-type flare trigger process is a ``reconnection-induced eruption" process where the tether-cutting reconnection of the flare is triggered before eruption of the flux rope.
It means that the RS-type flare does not necessarily cause a large flux rope eruption such as a CME.
%On the other hand, the OP-type is ``eruption-induced reconnection'' flare trigger process which should cause the flux rope eruption before the tether-cutting reconnection.
Therefore, it is also consistent that the flares triggered by the RS-type magnetic configuration did not have an associated CME in the AR.

\section{Summary} \label{sec:summary}

In this study, we analyzed the X1.0 and the preceding two C-class flares occurred in AR NOAA 12192.
We aimed to clarify the triggering process and to evaluate applicability of a flare trigger model such as proposed by \citet{kusano12}.
We analyzed the filter and vector magnetograms by Hinode/SOT and SDO/HMI, and also used the filtergrams for each layer of the solar atmosphere by SDO/AIA.
We found that there was a characteristic magnetic field structure that is the positive polarity intruding to the following negative sunspot (we named the structure ``the intruding positive polarity (IPP)'').
The significant chromospheric brightenings and the coronal loops were also observed around the IPP before the onset of the flares.
The relative shear angle $\chi$, that is defined as the angle between the potential field vector and the transverse field vectors, was measured in the AR.
These observed features were consistent with the RS-type flare trigger field of the KB12 model.
We considered the flare trigger scenario from the coronal magnetic loops that were extrapolated by the NLFFF method.
The triggering process of the flares was more consistent with the RS-type of the KB12 model rather than the case B of \citet{chen_shibata00}, because the former does not require epic eruption of flux rope trigger flare.
Therefore, we concluded that the RS-type of the KB12 model is more suitable for the triggering process of the C-class flares, and that the X1.0 flare was caused by rise of the flux rope that formed by the preceding C-class flares, in a three-step process.

In the AR, the RS-type field was located slightly off the main flaring PIL whereas the original RS-type simulation.
According to an extra simulation with an offset of the RS-type field from the highly sheared PIL (see the Appendix), it is suggested that the RS-type flare trigger process has certain flexibility for the distance of the flare trigger field from the highly sheared PIL.
%Moreover, we discussed consistency between the observed features and the RS-type flare trigger process when the trigger field locates slightly off the highly sheared PIL.
%We compared the observed brightenings and flare ribbons in AIA images to the current layers formed in the simulation in which the RS-type bipole field is injected off the PIL.
%The physical process of flare triggering was the same to the original RS-type scenario even in the case of the bipole field locates away from the PIL.
%However, the magnetic topology became much more complicated than the original RS-type.
%Even though, the locations of the intensive current layer expected in the simulation was consistent with the observed features in AIA images.
%This finding suggest that the RS-type flare trigger process has certain flexibility for the distance of the flare trigger field from the highly sheared PIL.
However, it is still unclear how a distant field that is away from the PIL can trigger flares on the actual solar surface.
We need to statistically investigate the distance between the flaring PILs and the flare trigger fields using observational data.

We also discussed the causality between these flares and CMEs by considering the critical height of the onset of the torus instability.
The closed magnetic field overlying a flux rope, which might be created by tether-cutting reconnection between the highly sheared loops, had not reached the critical decay index, and it might prohibit the flux rope from erupting.
In addition, this was also consistent with our conclusion that the C5.1, C9.7, and X1.0 flares were triggered by the RS-type field, which might cause reconnection without the help of eruptive instabilities such as the torus instability.
This suggests that the magnetic reconnection might not developed substantially, and that the flux rope did not reach a condition of the torus instability.

In the AR 12192, more five X-class flares occurred with the growth of the positive polarity, that is intruding to the following negative polarity region.
These X-class flares showed clear two-ribbon structures at similar locations, and all the X-class flares also did not associated with any CMEs.
We suggest that the west side of the IPP could be a trigger of these X-class flares although we need the detailed analysis for the each flare events.
\citet{sun15} reported that the electric current along the flaring PIL was very low, and horizontal magnetic field decreased much slower with height, before the X3.1 flare on October 24 ($\sim$ 20 hours before the X1.0 flare).
\citet{thalmann15} using global magnetic field modeling also pointed out that the strong closed arcade over the highly sheared arcades contributed to the confinement of the flux rope, for several M-class and X-class flares that occurred in AR 12192.
Therefore, a similar magnetic configuration might be kept during the AR disk passage, and it is important to investigate the temporal evolution of the characteristic magnetic structures in order to clarify the conditions of these ``consecutive'' and ``confined'' flares.
The combination of Hinode and SDO observations and NLFFF extrapolations is a powerful tool for analyzing local and global magnetic field changes in the AR.

\acknowledgments

%We thank the anonymous referee for improving the paper.
We are grateful to Dr. David H. Brooks and Hinode group members in ISAS/JAXA for useful discussions.
The HMI and AIA data have been used courtesy of NASA/SDO and the AIA and HMI science teams.
Hinode is a Japanese mission developed and launched by ISAS/JAXA, which collaborates with NAOJ as a domestic partner and with NASA and STFC (UK) as international partners.
Scientific operation of the Hinode mission is conducted by the Hinode science team organized at ISAS/JAXA.
This team mainly consists of scientists from institutes in the partner countries. Support for the post-launch operation is provided by JAXA and NAOJ (Japan), STFC (UK), NASA, ESA, and NSC (Norway).
This work was partly carried out at the NAOJ Hinode Science Center, which is supported by MEXT KAKENHI Grant Number 17GS0208, by generous donations from Sun Microsystems, and by NAOJ internal funding.
The Hinode Science Center at Nagoya University also supported the study.
Part of this work was carried out on the Solar Data Analysis System operated by the Astronomy Data Center in cooperation with the Hinode Science Center of the NAOJ.
%Also, we plotted the magnetic field lines extrapolated by NLFFF method using VAPOR, which is a product of the National Center for Atmospheric Research's Computational and Information Systems Lab.
This work was supported by JSPS KAKENHI Grant Numbers 23340045, 15H05814, and Grant-in-Aid for JSPS Fellows.

%\appendix
\section*{Appendix}

Figure~\ref{fig:offset} shows a snapshot of an extra simulation in which the RS-type bipole field is injected off the PIL.
White/black area indicates the positive/negative magnetic polarity area, and the green tubes represent the magnetic field lines.
The red surfaces correspond to the intensive electric current layers.
The major bipole is composed by the large positive (LP) and the large negative (LN) regions, and the major PIL is located between LP and LN.
The small-scale bipole field, that satisfies the RS-type condition, intrude into the LP region, and it is composed by the small positive (SP) and the small negative (SN) regions.
Note that the RS-type bipole field intrude to the positive polarity region while the IPP (positive polarity) intrude to the negative polarity (N1) region in the AR 12192.
Moreover, the magnetic helicity in the simulation is positive while the magnetic helicity in the analyzed active region is negative. 

\begin{figure*}
%\epsscale{2.00}
\plotone{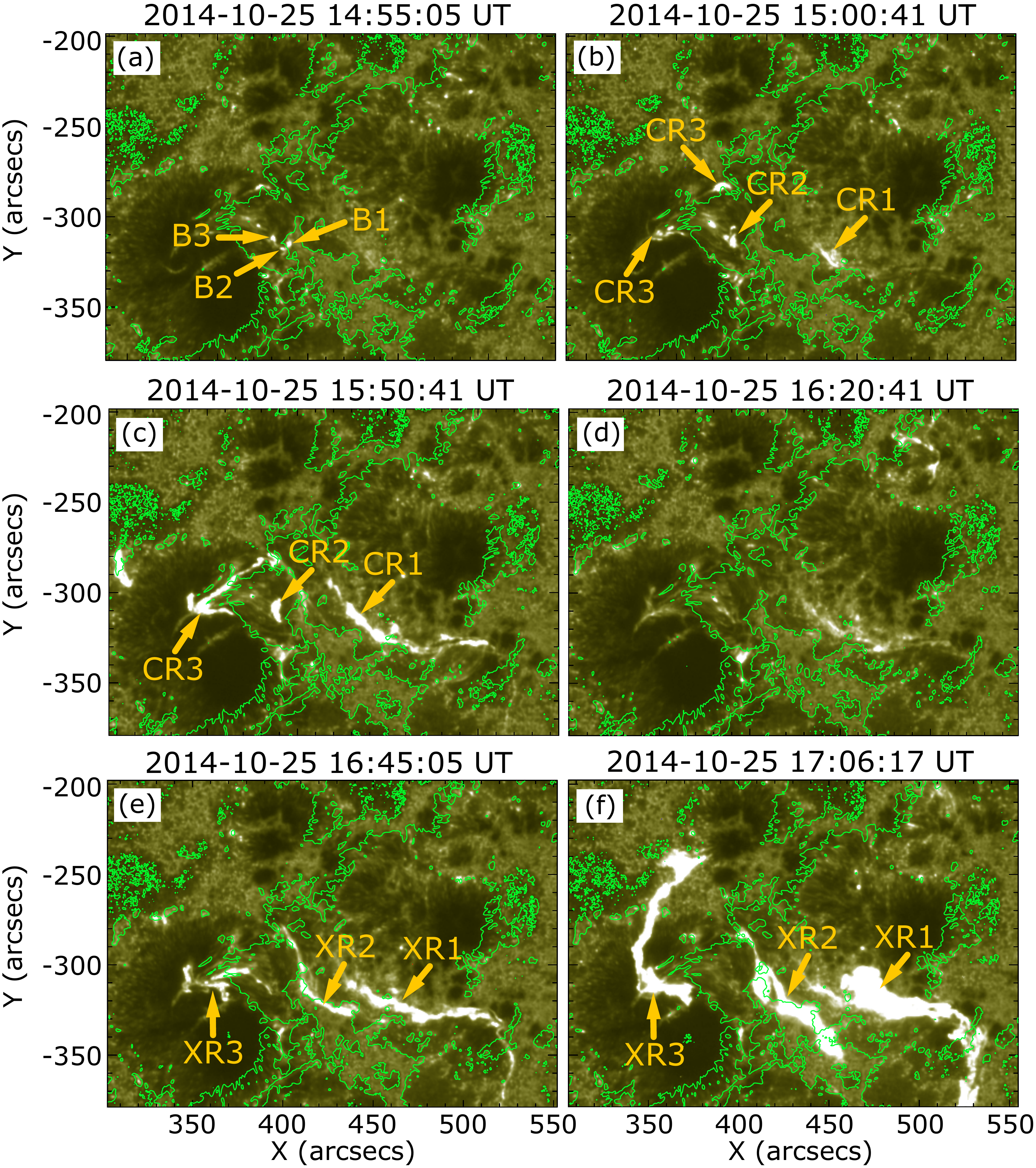}
\caption{
The temporal evolution of the significant brightenings in upper photosphere and the transition region, and the flare ribbons in AIA 1600 {\AA} images.
Green lines indicate the PILs in HMI LOS magnetograms at each time.
The intensity scale range is 0-2000 DN.
(a) Strong brightenings are intermittently seen at the west side of the IPP as indicated by the yellow arrow.
(b) The three-ribbons (CR1-CR3) of the C5.1 flare.
(c) The three-ribbons (CR1-CR3) of the C9.7 flare.
(d) A faint ribbon-like brightening remains at the region where the CR1 is seen.
(e) The initial three-ribbons of the X1.0 flare. XR1 is the positive ribbon, and XR2 and XR3 are the negative ribbons.
(f) Enhanced three-ribbon emission of the X1.0 flare.
}
\label{fig:AIA1600}
\end{figure*}

\begin{figure*}
%\epsscale{2.00}
\plotone{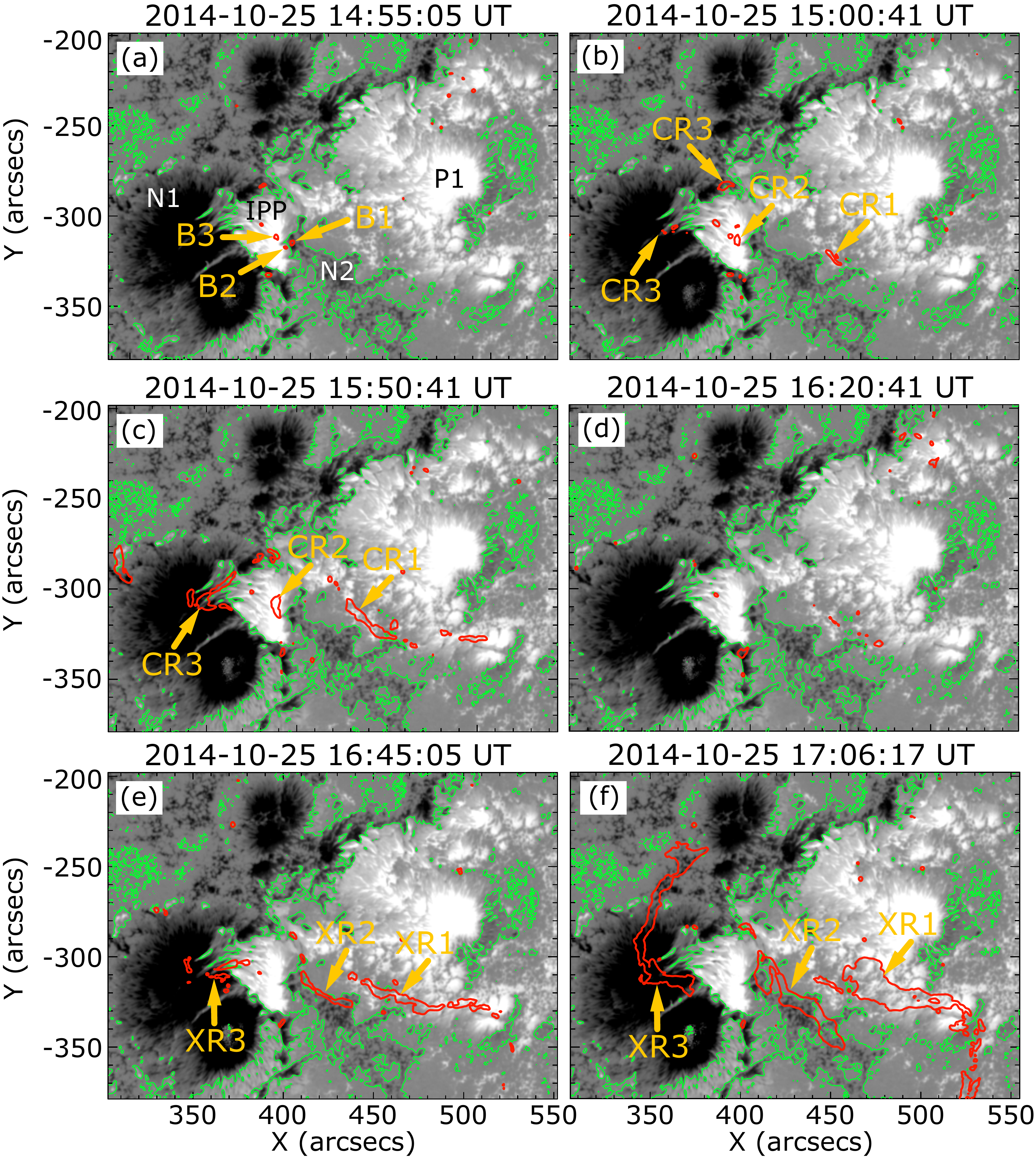}
\caption{
The temporal evolution of the HMI LOS magnetic field at $\pm1000$ G.
White/black indicates positive/negative polarity of the LOS magnetic field, and green lines indicate the PILs.
The red contours outline the brightenings in AIA 1600 {\AA} images with an intensity of 2000 DN.
The time of each panel is same as that of the AIA 1600 {\AA} images.
Panels (a)-(f) show the same features as Figure~\ref{fig:AIA1600} together the LOS magnetic field.
}
\label{fig:HMI1600}
\end{figure*}

\begin{figure*}
%\epsscale{2.00}
\plotone{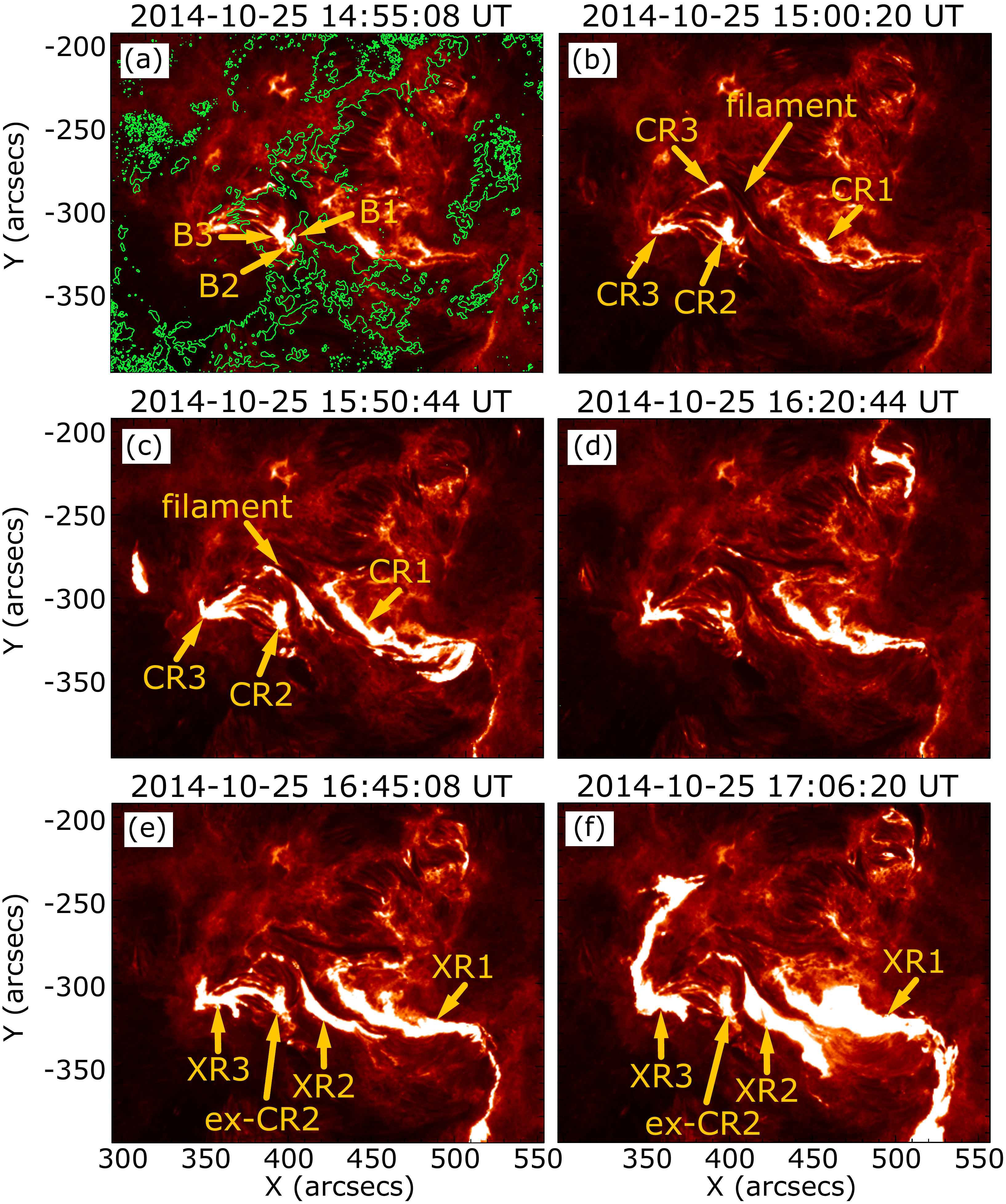}
\caption{
The temporal evolution of the chromospheric brightenings and flare ribbons in AIA 304 {\AA} images.
The PILs are over plotted with green lines on panel (a).
The intensity scale range is 0-1000 DN.
Panels (a)-(f) show almost the same features as Figure~\ref{fig:AIA1600}, but the brightening B1 shapes a small loop striding over the local PIL at the west side of the IPP.
Moreover, CR1 remains more clearly than that seen in Figure~\ref{fig:AIA1600}(c), and CR3 also remains in panel (d).
}
\label{fig:AIA304}
\end{figure*}

\begin{figure*}
%\epsscale{2.00}
\plotone{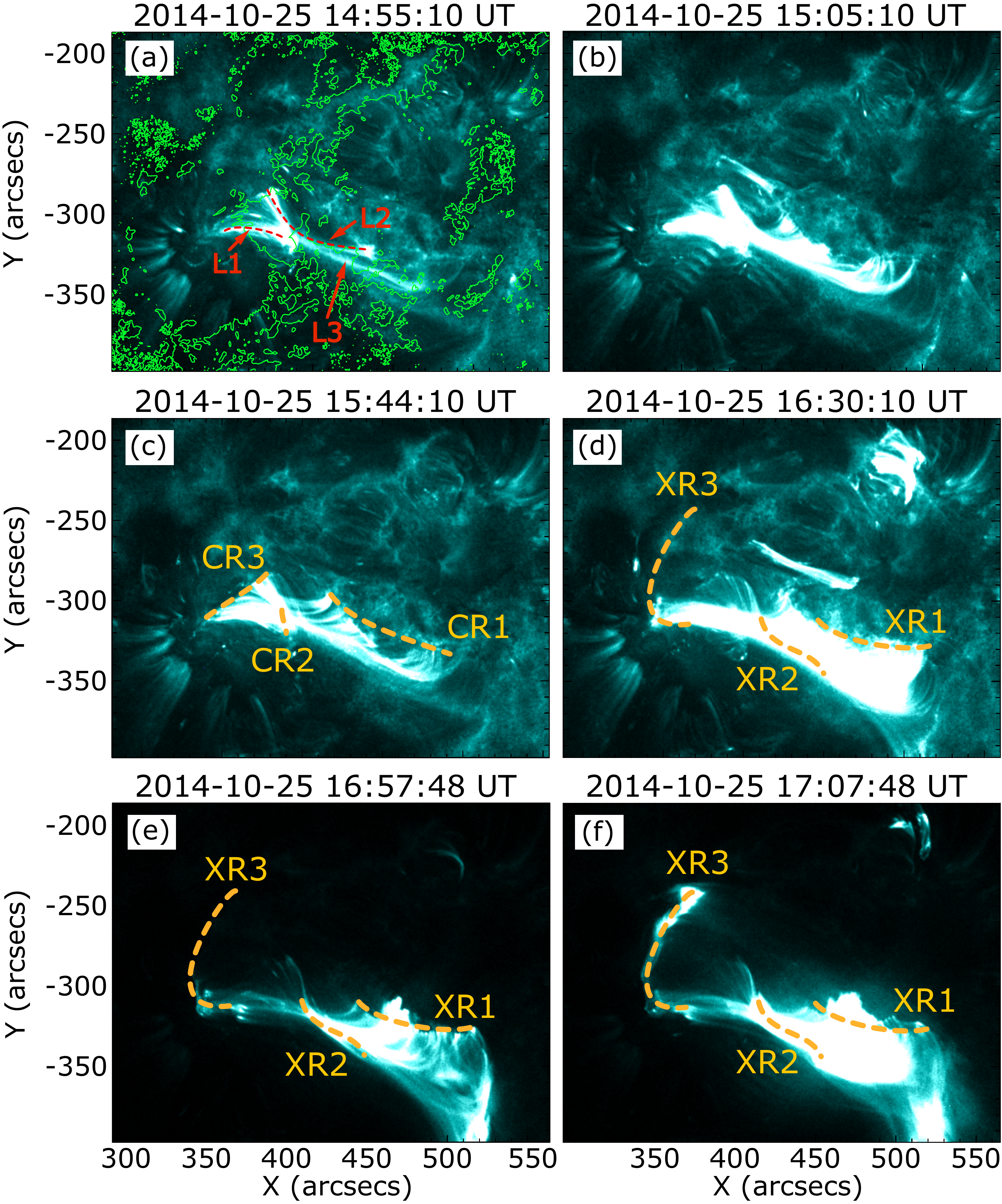}
\caption{
The temporal evolution of the coronal loops in AIA 131 {\AA} images.
The PILs are over plotted with green lines in panel (a).
The intensity scale range is 0-500 DN.
Bright loop L1 and L2, which connect IPP-N1 and P1-N1, are outlined by red broken lines in panel (a).
The other loop L3 likly connects P1-N2.
The yellow broken lines roughly illustrate the locations of the C9.7 three-ribbons in panel (c) and the locations of the X1.0 three-ribbons in panel (d)-(f).
}
\label{fig:AIA131}
\end{figure*}

\begin{figure*}
%\epsscale{2.00}
\plotone{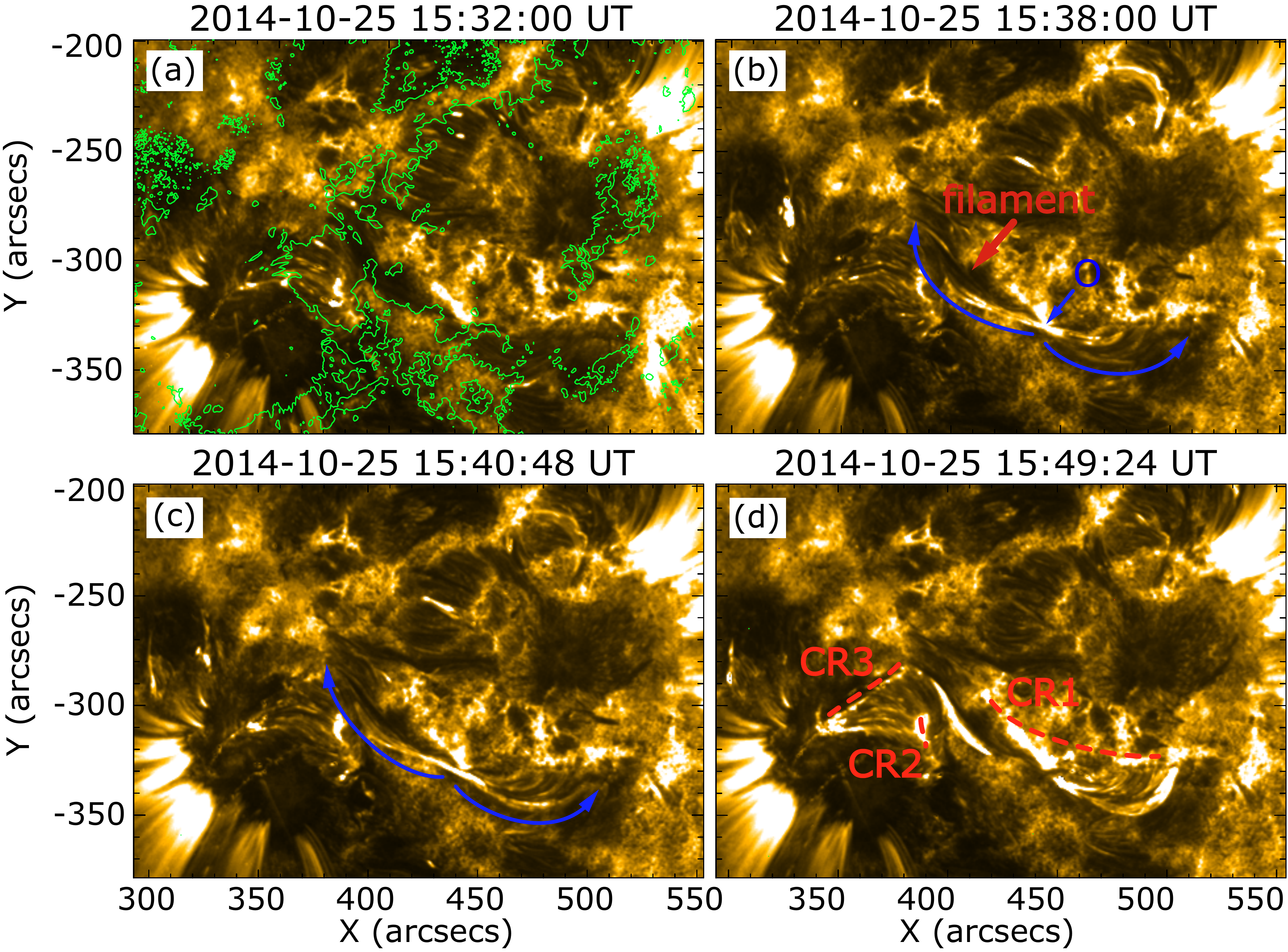}
\caption{
Tether-cutting magnetic reconnection in AIA 171 {\AA} images.
The FOV is almost the same as in Figures~\ref{fig:AIA1600}-\ref{fig:AIA131}.
The intensity scale range is 0-5000 DN, and the PILs are over plotted as the green lines in panel (a).
(b) Faint loop structures intersect at point O (indicated by the blue arrow), and small brightenings frequently move along the faint loops.
(c) The motion along the loop, such as illustrated by the blue arrows, are enhanced, and the loops slightly expand.
(d) CR1 and CR3 appear at the foot points of the faint loops as indicated by red broken lines. (Movie 1)
}
\label{fig:AIA171}
\end{figure*}

\begin{figure*}
%\epsscale{2.00}
\plotone{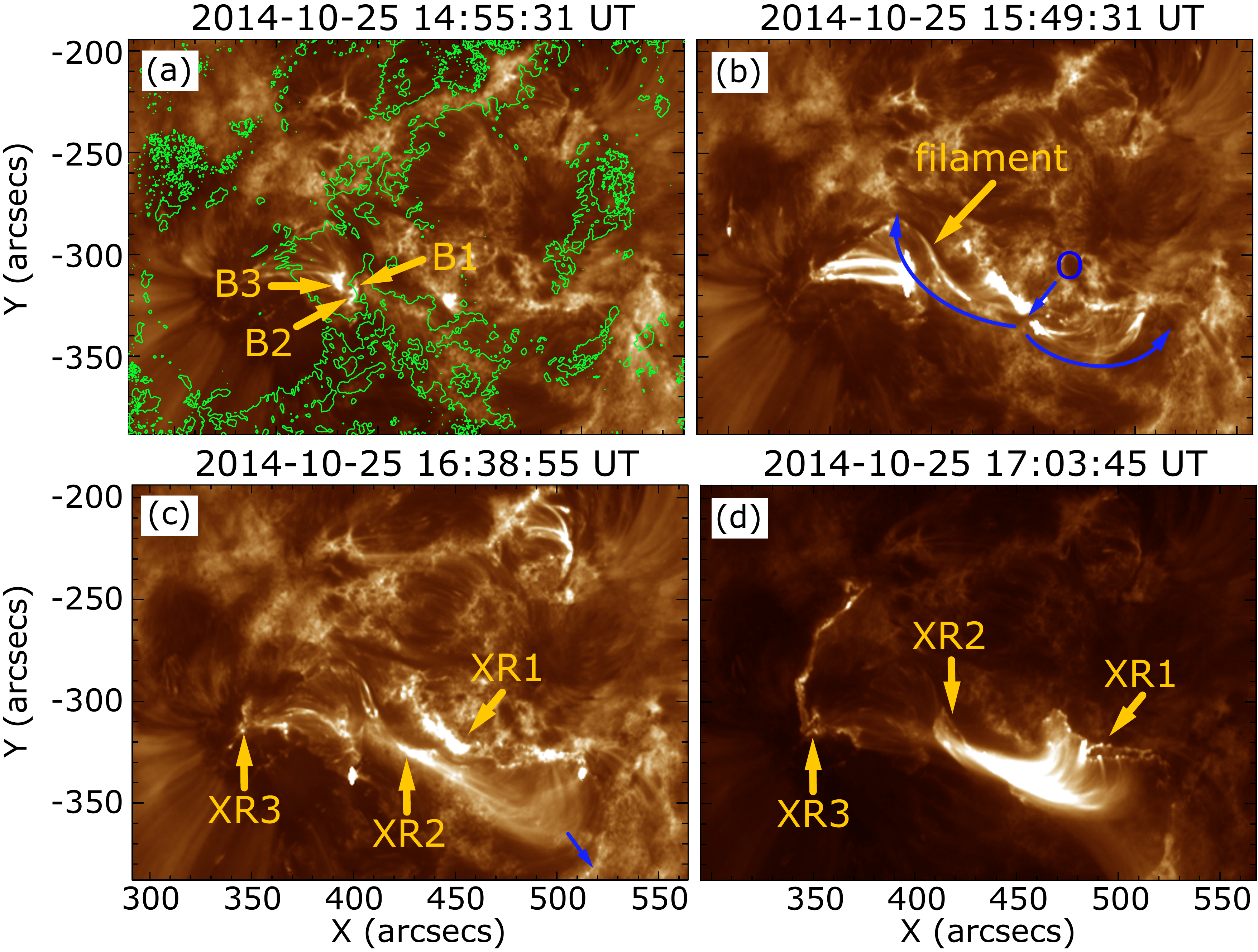}
\caption{
The evolution of the coronal loops and brightenings in AIA193 {\AA} images.
The FOV is almost the same as in Figures~\ref{fig:AIA1600}-\ref{fig:AIA131}.
The intensity scale range is 0-8000 DN, and the PILs are over plotted as the green lines in panel (a).
}
\label{fig:AIA193}
\end{figure*}

\begin{figure*}
%\epsscale{2.20}
\plotone{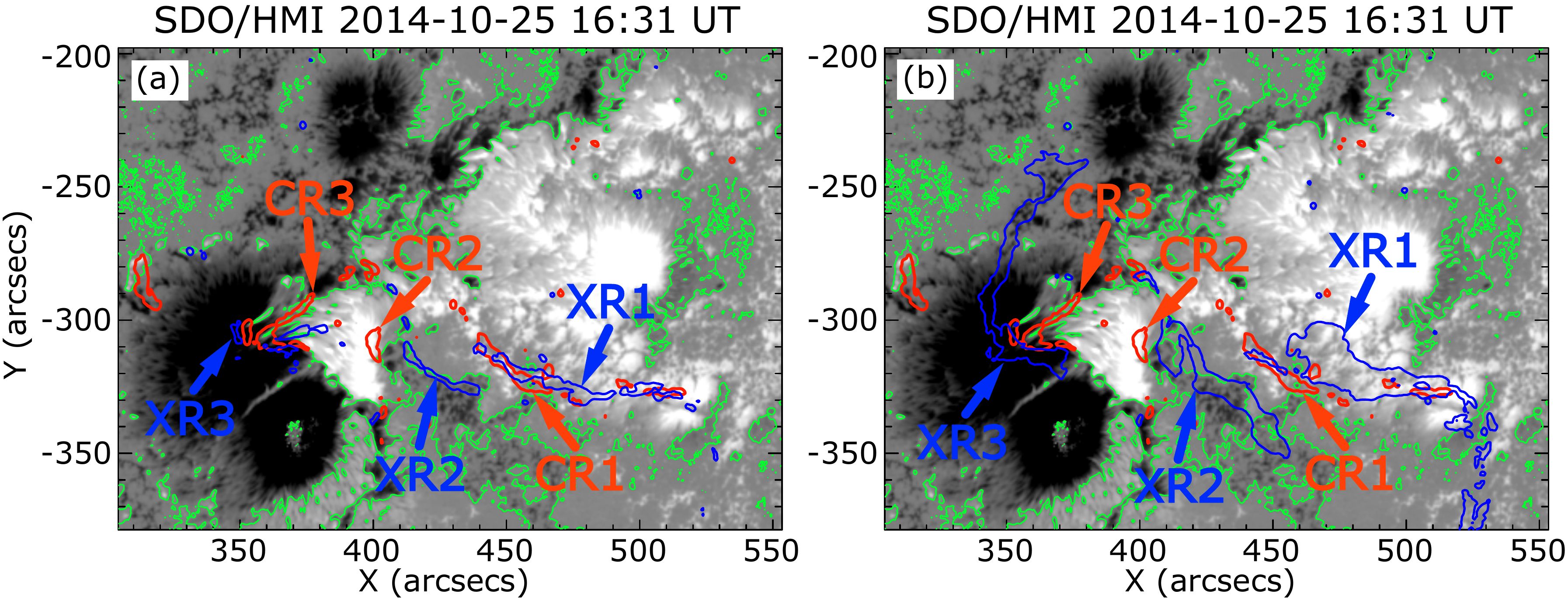}
\caption{
The flare ribbons of the C9.7 and the X1.0 flares.
The background images are HMI LOS magnetograms at the onset time of the X1.0 flare (16:31 UT on 2014 October 25).
The green lines are the PILs, and the red/blue contours outline the flare ribbons in AIA 1600 {\AA} images.
The red contour shows the CR1, CR2, and CR3 of the C9.7 flare at 15:50 UT in both panels (a) and (b).
The blue contour outlines the initial flare ribbons (XR1-XR3) of the X1.0 flare at 16:45 UT in panel (a), while the enhanced flare ribbons at 17:06 UT is outlined in panel (b).
}
\label{fig:flare_ribbons}
\end{figure*}

\begin{figure*}
%\epsscale{2.20}
\plotone{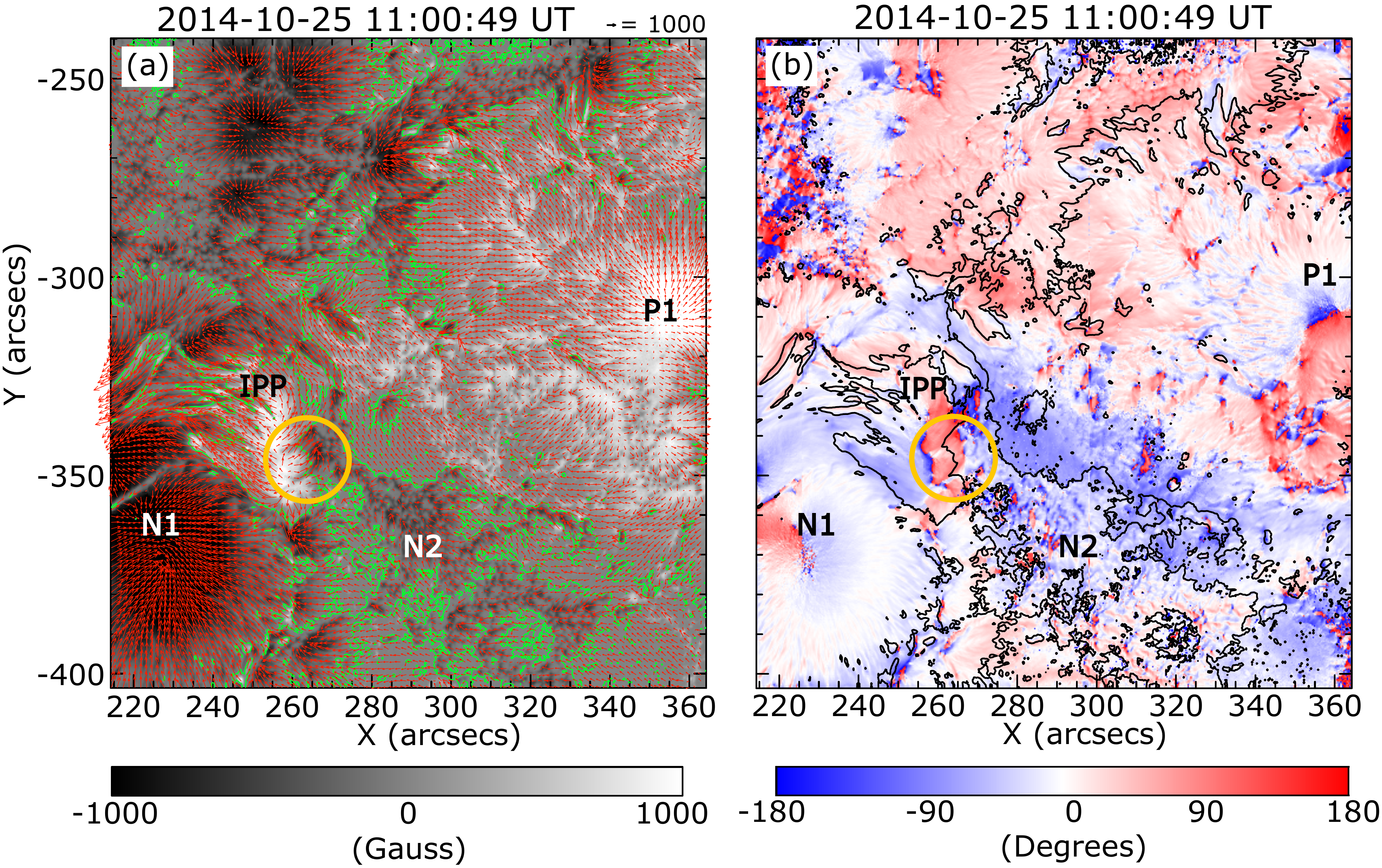}
\caption{
(a) The distribution of the heliographic magnetic field obtained by Hinode/SP between 11:00 to 11:33 UT on 2014 October 25.
The grayscale part corresponds to the positive/negative polarity of $B_{z}$ at $\pm2000$ G.
Green lines indicate the PILs (line of $B_{z} = 0 G$), and the red arrows are the vectors of the horizontal magnetic field $\bm B_{h} = \sqrt{{B_{x}}^2 + {B_{y}}^2}$ at each point.
(b) Distribution of the relative shear angle $\chi$ which is defined as the angle between the potential field vector ${\bm B_{p}}$ and horizontal field vector ${\bm B_{h}}$.
The black lines are the PILs which are indicated by green lines in panel (a).
Red/blue corresponds to positive/negative values of $\chi$, i.e. the magnetic helicity.
The region where the strong chromospheric emissions were seen is pointed out by the yellow circle in both (a) and (b).
}
\label{fig:SP}
\end{figure*}

\begin{figure*}
%\epsscale{2.20}
\plotone{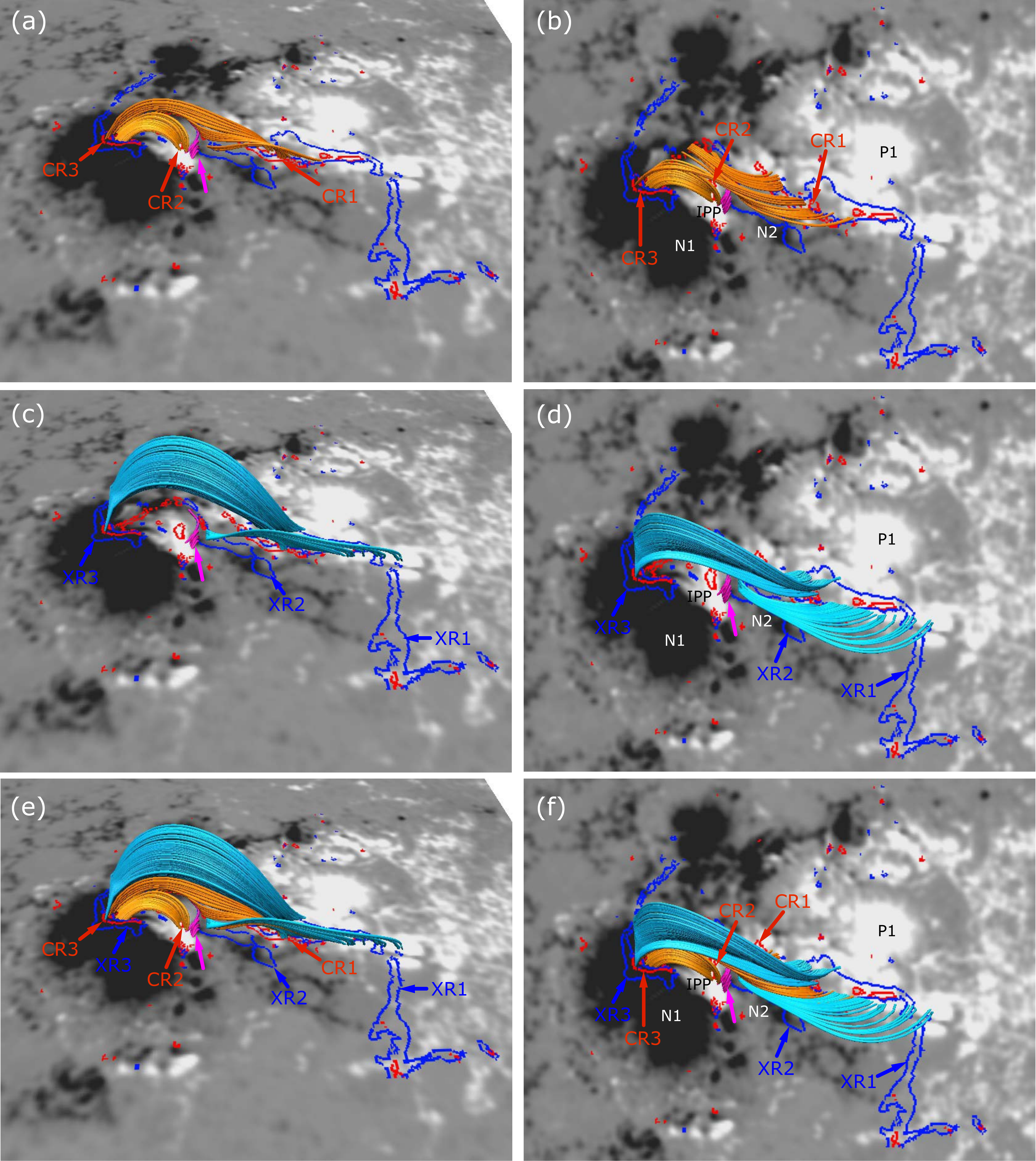}
\caption{
The magnetic field lines extrapolated by the NLFFF method.
The left/right columns show a bird's eye view/top view of the same image.
The grayscale images are the HMI LOS magnetogram taken at October 25 15:00 UT.
The red/blue contour outlines the brightenings (700 DN) in AIA 1600 {\AA} such as flare ribbons on 15:50 UT/17:03 UT.
The small magenta tubes indicated by the magenta arrows are the local magnetic field lines at the west side of the IPP.
The orange/sky blue tubes indicate the coronal magnetic field lines anchor to the flare ribbons of the C9.7/X1.0 flare in panels (a) and (b)/(c) and (d), respectively.
All the coronal magnetic field lines are plotted in panels (e) and (f).
}
\label{fig:nlfff}
\end{figure*}

\begin{figure*}
%\epsscale{2.00}
\plotone{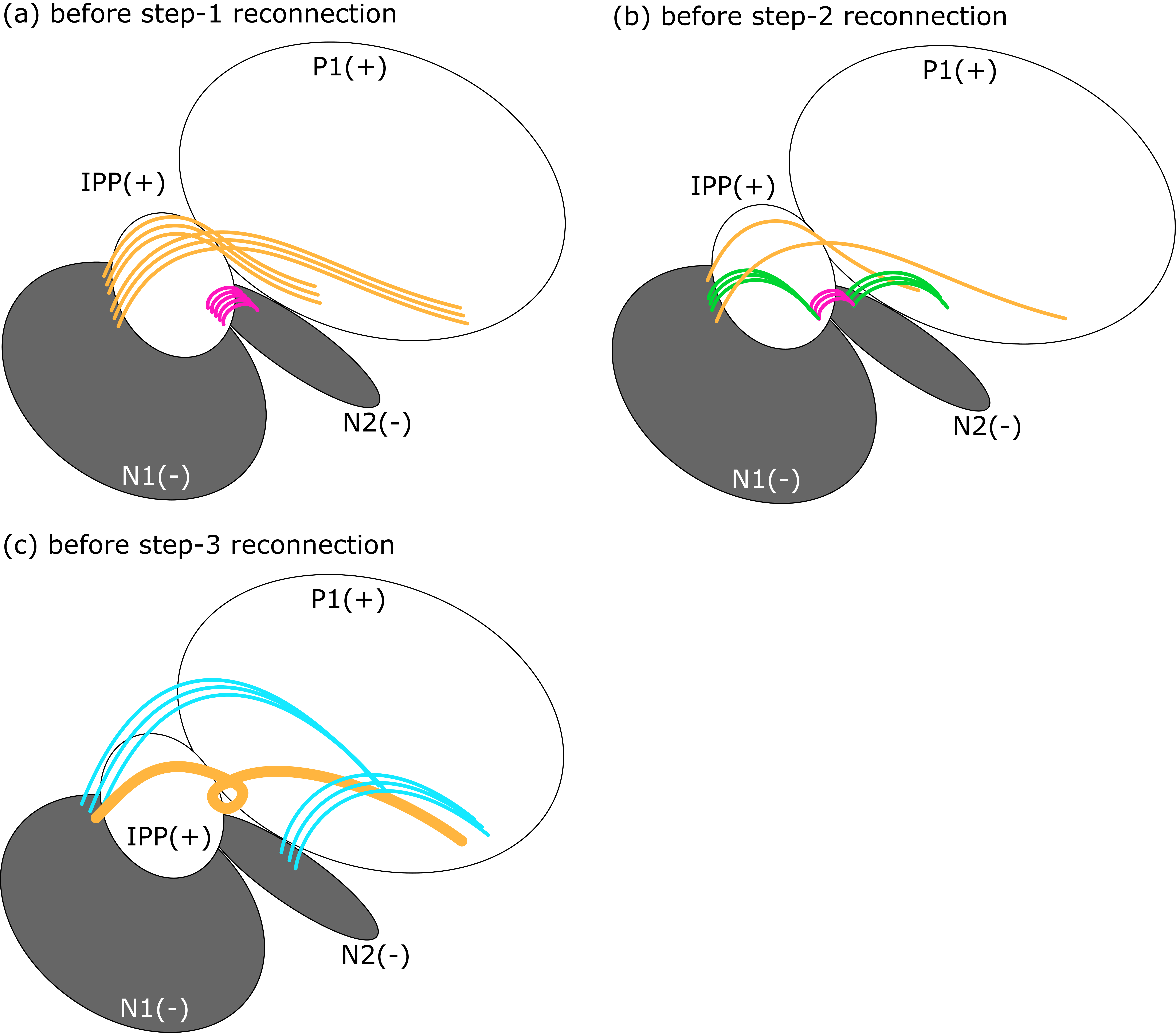}
\caption{
The schematics of the magnetic field lines before step-1, step-2, and step-3 reconnection.
White/gray indicates positive/negative polarity of the LOS magnetic field.
(a) Before step-1 reconnection. The orange and magenta loops illustrate the P1-N1 and IPP-N2 loops in Figure~\ref{fig:nlfff} (a, b).
(b) Before step-2 reconnection (i.e. after step-1 reconnection). The orange and magenta loops are the same to these in panel (a). The green loops indicate the small loops connecting the IPP-N1 and P1-N2 that are formed by the flux cancellation between the orange and magenta loops.
(c) Before step-3 reconnection. The sky blue loops are equivalent to the P1-N2 and P1-N1 loops in Figure~\ref{fig:nlfff} (c, d). The thick orange line illustrates the flux rope that is formed by step-2 reconnection under the sky blue loops.
}
\label{fig:schematic}
\end{figure*}

\begin{figure*}
%\epsscale{2.20}
\plotone{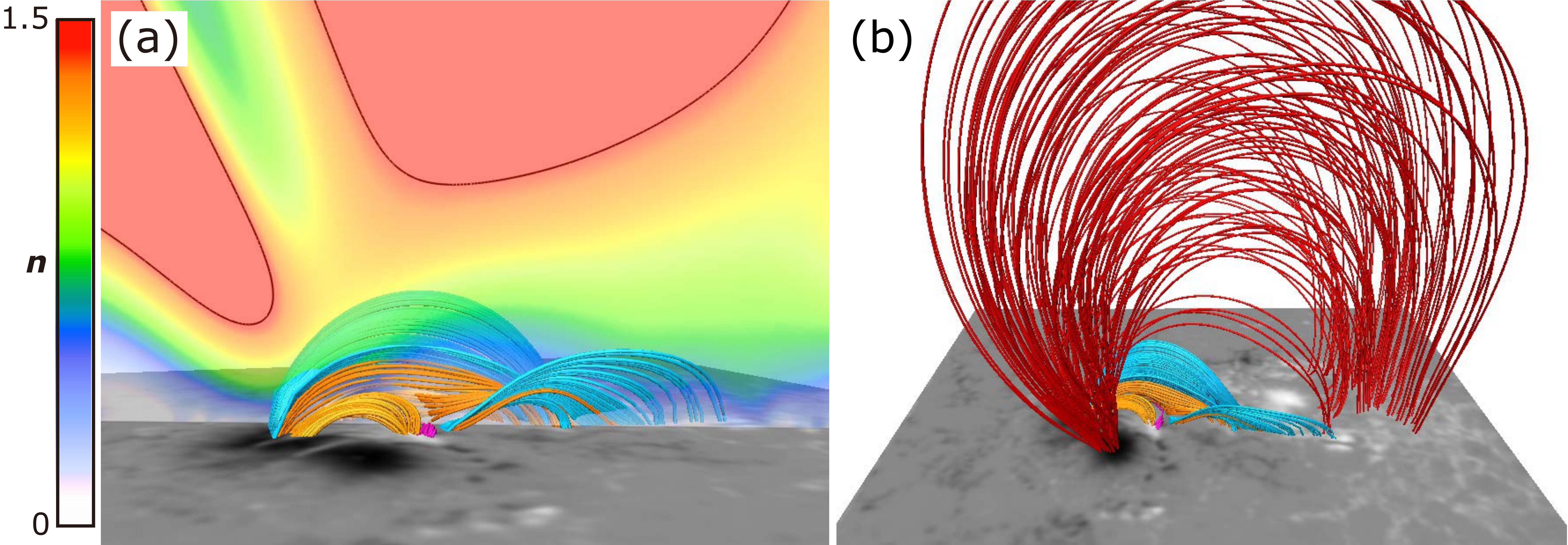}
\caption{
The coronal magnetic field lines from a side view of Figure~\ref{fig:nlfff}(f).
The vertical cross section in panel (a) displays the decay index {\it n} between 0 to 1.5.
The decay index {\it n} is calculated form the horizontal component of the potential field extrapolated from the vector magnetic field.
The back line on the vertical cross section indicates {\it n} = 1.5 which corresponds to the critical value to appear the torus instability.
The red loops in panel (b) are closed loops over the highly sheared loops (the orange and the sky blue loops).
Obviously, the highly sheared loops corresponding to the C9.7 and the X1.0 flare are lower than the isosurface of {\it n} = 1.5, and there are closed loops over the sheared loops.
}
\label{fig:DI1p5}
\end{figure*}

\begin{figure*}
%\epsscale{2.00}
\plotone{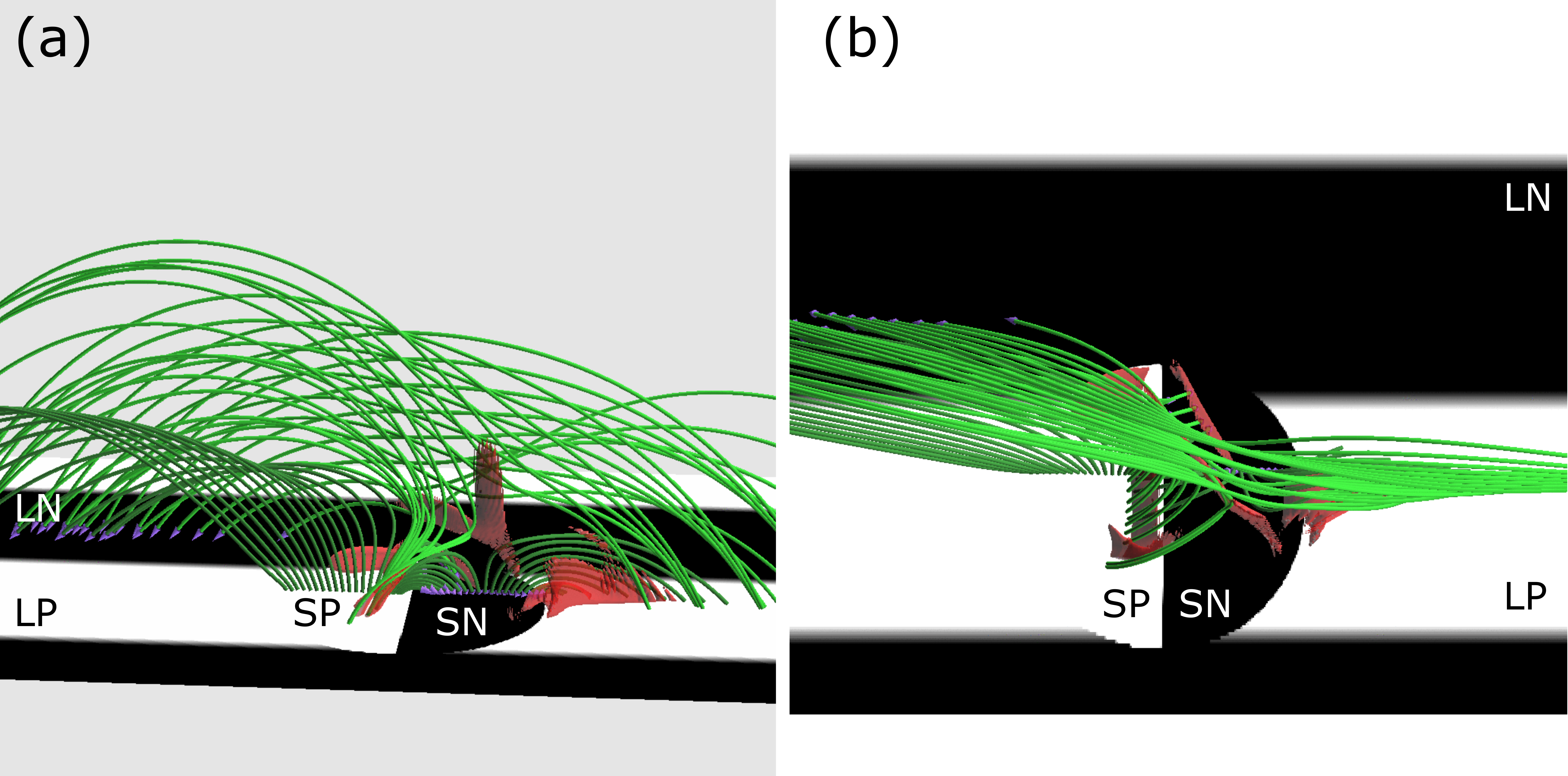}
\caption{
A snapshot of the KB12 model simulation with an offset of the RS-type field from the PIL.
White and black areas indicate the positive and negative magnetic polarity areas, respectively.
Green tubes represent the magnetic field lines, and red surfaces correspond to the intensive current layers.
Both (a) and (b) show the same snapshot, but the bird's eye view and top view are shown in panel (a) and (b), respectively.
}
\label{fig:offset}
\end{figure*}

Figure~\ref{fig:offset} shows the snapshot between the step-1 and step-2 reconnection, i.e. between the internal reconnection and flare reconnection of the RS scenario in KB12 model.
The overlying field connecting LP-LN and small bipole field SP-SN are equivalent to the P1-N1 loops (orange loops) and the IPP-N2 loops (magenta loops) in Figures~\ref{fig:nlfff} and \ref{fig:schematic}, respectively.
The small loops connecting SP-LN and LP-SN are equivalent to the IPP-N1 and P1-N2 (green loops in Figure~\ref{fig:schematic}(b)).
The vertical electric current layer is formed over the center of the SN region as the step-1 reconnection proceeds between the LP-LN and SP-SN loops.
This vertical electric current is observed as the brightening B3 and CR2 in AIA 1600 {\AA} and 304 {\AA} images (Figures~\ref{fig:AIA1600} and \ref{fig:AIA304}).
Then the large-scale sheared loops (LP-LN) collapse inward and the flare ribbons appeared at the foot points in the LP and LN regions.
These ribbons are equivalent to the CR1 and CR3 of the C-class flares.
Therefore, the locations of the intensive electric current layer are consistent with the RS-type case.

Moreover, the twisted flux rope, which is formed by the reconnection between LP-LN and SP-SN loops, asymmetrically rise up from the LN-side in the simulation.
In the analyzed active region, the flux rope illustrated by the thick orange line in Figure~\ref{fig:schematic} starts eruption from the P1-side.
Therefore, the step-3 reconnection of the sky-blue loops connecting P1-N2 occurs earlier by the asymmetric rise of the underlying flux rope, and the flare ribbons XR1 and XR2 appears earlier than XR3.

\clearpage

\begin{table*}
\begin{center}
\caption{A list of X-class flares that occurred in NOAA AR 12192}
\begin{tabular}{cccccc}
\tableline\tableline
date & start time\tablenotemark{a} & peak time\tablenotemark{a} & end time\tablenotemark{a}
& GOES X-ray class & location\tablenotemark{b} \\
& (UT) & (UT) & (UT) & & \\
\tableline
19 October 2014 & 04:17 & 05:03 & 05:48 & X1.1 & S14E64 \\
22 October 2014 & 14:02 & 14:28 & 14:50 & X1.6 & S14E13 \\
24 October 2014 & 21:07 & 21:40 & 22:13 & X3.1 & S22W21 \\
25 October 2014 & 16:31\tablenotemark{c} & 17:03\tablenotemark{c} & 17:07\tablenotemark{c} & X1.0 & S10W22 \\
26 October 2014 & 10:04 & 10:56 & 11:18 & X2.0 & S14W37 \\
27 October 2014 & 14:02 & 14:47 & 15:09 & X2.0 & S16W56 \\
\tableline
\label{table:eventlist}
\end{tabular}
%% Any table notes must follow the \end{tabular} command.
\tablenotetext{a}{The start, peak, and end times are defined from X-ray observations by the GOES satellite.}
\tablenotetext{b}{\url{http://www.solarmonitor.org/}}
\tablenotetext{c}{These times are defined from RHESSI observations.}
%\tablecomments{@@@@@}
\end{center}
\end{table*}

%\begin{table}
%\begin{center}
%\caption{Summary of SDO/AIA wavelength\tablenotemark{c}}
%\begin{tabular}{llll}
%\tableline\tableline
%wavelength & primary ion(s) & Region of atmosphere & Temperture  \\
%& & & (log (T)) \\
%\tableline
%304{\AA} & \ion{He}{2} & chromosphere, transition region & 4.7 \\
%1600{\AA} & \ion{C}{4} & upper photosphere, transition region & 5.0 \\
%171{\AA} & \ion{Fe}{9} & upper transition region, quiet corona & 5.8 \\
%131{\AA} & \ion{Fe}{8}, \ion{Fe}{21} & transition region, flaring corona & 5.6, 7.0 \\
%\tableline
%\label{table:aia_lines}
%\end{tabular}
%% Any table notes must follow the \end{tabular} command.
%\tablenotetext{c}{\citep[The information in this table are excerpts from ][]{lemen12}}
%%\tablecomments{@@@@@}
%\end{center}
%\end{table}

\end{document}